\documentclass[prodmode]{acmsmall} 

\usepackage[ruled]{algorithm2e}
\usepackage[T1]{fontenc}
\usepackage[latin9]{inputenc}
\usepackage{array}
\usepackage{float}
\usepackage{mathrsfs}
\usepackage{multirow}
\usepackage{amsmath}
\usepackage{amssymb}
\usepackage{graphicx}
\usepackage{listings}
\usepackage{xcolor}

\SetAlFnt{\small}
\SetAlCapFnt{\small}
\SetAlCapNameFnt{\small}
\SetAlCapHSkip{0pt}
\IncMargin{-\parindent}
\providecommand{\tabularnewline}{\\}
\floatstyle{ruled}
\newfloat{algorithm}{tbp}{loa}
\providecommand{\algorithmname}{Algorithm}
\floatname{algorithm}{\protect\algorithmname}

\lstdefinelanguage[OpenCL]{C}[ANSI]{C} 
{morekeywords={__kernel,kernel,__local,local,__global,global,%
__constant,constant,__private,private,%
char2,char3,char4,char8,char16,%
uchar2,uchar3,uchar4,uchar8,uchar16,%
short2,short3,short4,short8,short16,%
ushort2,ushort3,ushort4,ushort8,ushort16,%
int2,int3,int4,int8,int16,%
uint2,uint3,uint4,uint8,uint16,%
long2,long3,long4,long8,long16,%
ulong2,ulong3,ulong4,ulong8,ulong16,%
float2,float3,float4,float8,float16,%
image2d_t,image3d_t,sampler_t,event_t,%
bool2,bool3,bool4,bool8,bool16,%
half2,half3,half4,half8,half16,%
quad,quad2,quad3,quad4,quad8,quad16,%
complex,imaginary},%
}%

\begin{document}

\markboth{H. Wang et al.}{FPGA-based Acceleration of FT Convolution for Pulsar Search Using OpenCL}

\title{FPGA-based Acceleration of FT Convolution for Pulsar Search Using OpenCL}
\author{HAOMIAO WANG
PRABU THIAGARAJ
OLIVER SINNEN
}

\begin{abstract}
The Square Kilometre Array (SKA) project will be the world largest
radio telescope array. With its large number of antennas, the number
of signals that need to be processed is dramatic. One important element
of the SKA's Central Signal Processor package is pulsar search. This
paper focuses on the FPGA-based acceleration of the Frequency-Domain
Acceleration Search module, which is a part of SKA pulsar search engine.
In this module, the frequency-domain input signals have to be processed
by 85 Finite Impulse response (FIR) filters within a short period
of limitation and for thousands of input arrays. Because of the large
scale of the input length and FIR filter size, even high-end FPGA
devices cannot parallelise the task completely. We start by investigating
both time-domain FIR filter (TDFIR) and frequency-domain FIR filter
(FDFIR) to tackle this task. We applied the overlap-add algorithm
to split the coefficient array of TDFIR and the overlap-save algorithm
to split the input signals of FDFIR. To achieve fast prototyping design,
we employed OpenCL, which is a high-level FPGA development technique.
The performance and power consumption are evaluated using multiple
FPGA devices simultaneously and compared with GPU results, which is
achieved by porting FPGA-based OpenCL kernels. The experimental evaluation
shows that the FDFIR solution is very competitive in terms of performance,
with a clear energy consumption advantage over the GPU solution.
\end{abstract}

%
%


%
%




\begin{bottomstuff}
This work is supported by the Parallel and Reconfigurable Computing Lab and 
SKA New Zealand Alliance under the SKA Project Scholarship.

Author's addresses: H. Wang, Department of Electrical and Computer Engineering,
University of Auckland, Private Bag 92019, Auckland 1142,
New Zealand; Email: hwan938@auckland.ac.nz; {and} O. Sinnen, 
Department of Electrical and Computer Engineering,
University of Auckland, Private Bag 92019, Auckland 1142,
New Zealand; Email: o.sinnen@auckland.ac.nz.
\end{bottomstuff}

\maketitle

\section{Introduction}

Pulsar (Pulsating Radio Source), as a highly magnetized rotating neutron
star, is an ideal research object for physics and astrophysics research.
It has been used in a wide range of areas, such as tests of general
relativity, galactic studies, and cosmology~\cite{carilli2004science}.
Unlike other visible astronomical objects, electromagnetic radiation
beams emitted from pulsars are hard for the optical telescope to detect
unless the beam is towards telescope and there are no obstacles between
them. 

Hence, employing radio telescopes becomes the main approach to observe
pulsars, and, in practice, most known pulsars have been recorded by
radio telescopes, including the first detected pulsar. The pulsar
signals are weak radio sources and some are even weaker than thermal
noise. So the physical scale of a radio telescope has to be large,
and the integration time of a specific space area has to be long enough.
For basic pulsar search, the input signals are Fourier transformed
and analysed in the frequency domain, due to the regularity of pulsar
beams. Since some pulsar beams might be scattered by clumpy interstellar
medium, a group of dispersion measure (DM) trails of one input array
has to be tested.

If the period of a pulsar is not a constant during the integration
period such as in binary pulsars (which science is most interested
in), the Doppler effect makes it more difficult to be observed. For
such pulsars, acceleration search is applied by assuming a group of
accelerations of a pulsar. This paper focuses on the Fourier Domain
Acceleration Search (FDAS). It is an effective approach to remove
(sweep up) the smearing of signals by using the correlation technique~\cite{ransom2002fourier}\cite{jouteux2002searching}. 

The Square Kilometer Array (SKA) \footnote{www.skatelescope.org}
will be the world's largest radio telescope array and is currently
in its phase one (SKA1) pre-construction. Based on the covered frequency
bandwidth, it is divided into LOW, MID, and SURVEY. Because of the
huge size and number of antennas, the workload of digital signal processing
becomes the main challenge for the hardware development~\cite{dewdney2009square}.
The Central Signal Processor (CSP) package of the SKA1-MID contains
many sub-elements, and the pulsar search engine (PSS) is one of them.
The PSS searches for pulsars over a range of dispersion measure, acceleration,
and period search space using various approaches. The FDAS module,
which is a compute-intensive application, is an essential part of
the SKA1-MID PSS. The core computation part of the FDAS module is
to convolve with a large number of input signals with a group of lengthy
templates. Because of the need for the precision, the input signals
and coefficients arrays are complex single-precision floating-point
data. The large amount of complex floating-point operations and restrict
time limitation are big problems for processors. Even for the powerful
high-performance computing systems, the overall workload is very challenging,
especially its very high power consumption, considering the remote
locations of the telescopes. 

It is essential for the SKA project to employ high-performance computing
devices in accelerating the compute-intensive signal processing tasks.
For such a large-scale project, it is necessary to consider all alternatives
before deployment. To evaluate the performance regarding execution
latency and energy dissipation for different hardware platforms, employing
the Field-programmable gate array (FPGA) for prototype design is necessary.
However, the traditional hardware development flow using Hardware
Description Languages (HDLs) such as VHDL or Verilog is a large barrier
to proper software engineering. It excludes non-hardware experts from
participating and following the development. Solutions often become
very device specific. This is especially problematic in the SKA project,
where the final hardware architecture and devices (e.g., FPGA versus
GPU) are not finalized yet. As an international research project,
many teams are working worldwide on different aspects, and high-level
approaches can strongly aid the interaction between different research
teams, especially for prototype designs. 

In this paper, we are investigating the use of FPGAs for the efficient
high-performance computing of the core computation part of the FDAS
module\textendash FT convolution. This not only produces low-power
processing solutions for the demanding pulsar search modules, but
also evaluates the viability of using high-level approaches to achieve
the needed efficiency and performance, and its ability to support
the sweeping of a large design space. The main contributions are as
follows:
\begin{itemize}
\item Investigation and proposal of various differing designs for FT convolution.
In contrast to previous work the designs are tailored to the demanding
nature of the underlying FIR filters: large filters (i.e., a large
number of coefficients), multiple filters for same input data, long
input data stream, complex floating point values and demanding real
time limit. 
\item Exploration of the design space in several directions: convolution
methods (time versus frequency based), area versus time efficient
designs, optimising for single filter versus multiple filters, FFT
parameters, single and multiple FPGAs, etc.
\item Implementation of designs using high-level approach OpenCL and analysis
of achieved performance relative to upper bounds (based on available
resources).
\item Extensive experimental evaluation of all designs on the real host
system with up to three FPGA boards; porting of OpenCL implementation
to GPU; performance comparisons regarding speed (i.e., execution latency)
and energy consumption. 
\end{itemize}
This paper studies the FPGA-based acceleration of the FT convolution
part of the FDAS module using high-level synthesis approach, and it
is organized as follows. In Section~\ref{Background}, the one-dimensional
convolution that is the key element of the FT convolution and hardware
acceleration in the radio astronomy area are discussed. We introduce
the basic FDAS module and time-domain and frequency-domain based algorithms
to handle its FT (Fourier Transform) convolution part in Section~\ref{FDAS-Module}.
The FPGA-based FPGA-based FIR filter structure and optimisation for
FT convolution module are proposed in Section~\ref{FIR_Structure}.
The FPGA-based OpenCL development technique is mentioned in Section~\ref{Architecture-and-Optimisation},
alongside a discussion of the portability of the FPGA-based kernels
to other platforms. In Section~\ref{Evaluation}, the performance
of a group of FIR filter implementations are evaluated, and the fastest
design is used to compared with GPU-based kernels. The conclusions
are drawn in Section~\ref{Conclusion}.

\section{Related Work}\label{Background}

\subsection{One-dimensional Convolution}

In the FT convolution module, the compute-intensive part is a large
number of lengthy FIR filters, whose essence is one-dimensional (1D)
convolution. The 1D convolution and FPGA-based acceleration of 1D
convolution have been well researched. 

Both Intel and Xilinx (time domain transpose direct form) provide
FIR compiler. For the non-symmetry coefficient array, they employ
the transpose multiply-accumulate architecture that implements the
FIR filter in time-domain. The SPIRAL project provides a multiplierless
FIR/IIR generator that uses only additions/subtractions and shifts
instead of multiplications. However, the data types of input signals
and coefficient array have to be fixed-point and it supports maximum
10-tap FIR filter (as opposed to several hundreds in this work). 

A thorough investigation of 1D convolution across different platforms
is done in~\cite{fowers2013performance}. The 1D convolution is implemented
in both time-domain using the overlap-save algorithm and frequency-domain
using the overlap-save algorithm. The evaluation showed that when
the template size is several hundred, the standalone frequency-domain
FPGA performs faster than GPUs and multicore processors. The data
types of the input signals and coefficient array are single-precision
floating points. Regarding the optimisation of a group of 1D convolutions
with complex single-precision floating-point operations, we believe
that it has not been addressed in literature before.

\subsection{Accelerator in Radio Astronomy}

High-end FPGAs, as accelerators, are widely employed in accelerating
large-scale computation such as Microsoft's hyperscale datacenters~\cite{putnam2014reconfigurable}
and IBM's Supervessel Cloud~\cite{chen2014enabling}. In many radio
astronomy projects, FPGA accelerators are employed to handle large-scale
computation as well. In~\cite{de2007radio}, hundreds of Xilinx Virtex-4
FPGAs are used to implement the correlator of the SKAMP project. In
LOFAR~\cite{van2009using,van2011correlating}, multi-core CPUs and
many-core architectures are evaluated to implement the correlator,
however, the power consumption is a problem. The Berkeley CASPER group,
MeerKAT, and NRAO released the FPGA-based acceleration hardware for
implementing the FX correlator of a radio telescope array~\cite{parsons2009digital}.
In~\cite{sanchez2005digital}, FPGA platforms are employed to handle
digital channelised receivers. On the GPU side, the NVIDIA GTX480
GPU based cross-correlation implementation for radio astronomy is
evaluated in~\cite{clark2012accelerating}. Although each device
could achieve over $1TFLOPS$ performance, the thermal design power
(TDP) of each device is over $250W$. 

\subsection{High-level Synthesis Approaches}

To reduce the time-to-market and increase the portability of source
code for FPGAs, a number of research works regarding high-level development
techniques have been undertaken~\cite{edwards2006challenges}. Known
examples are LegUp~\cite{canis2011legup,canis2013legup}, ROCCC~\cite{najjar2013fpga}
and Nimble~\cite{li2000hardware}, all of them can compile C based
code and generate bitstream files to program FPGAs. LegUp is an open
source project that is available for researchers to use. In terms
of commercial tools, Maxeler provides a compiler for its FPGA products
that are based on both Java and C/C++. LabVIEW~\cite{andrade1998software,kehtarnavaz2010digital}
provides a graphical programming environment to develop for its Xilinx
FPGA-based devices.

Besides these, the two primary FPGA vendors, Intel and Xilinx, provide
high-level development tools as well. The Vivado HLS, which is based
on AutoESL~\cite{zhang2008autopilot}, is widely adopted for high-level
Xilinx FPGA development, targeting C, C++ and System C~\cite{schmid2016big}.
Another high-level development environment is Xilinx's SDAccel \cite{wirbel2014xilinx},
which is designed for OpenCL applications targeting Xilinx FPGA-based
accelerator cards~\cite{guidi2016improve}\cite{guidi2016improve,fifield2016optimizing}.
Intel released a development tool called Altera SDK for OpenCL~\cite{chen2012invited,chen2013fractal,czajkowski2012opencl_1,czajkowski2012opencl},
which is based on OpenCL standard version 1.0. This OpenCL approach
seems very promising as it is not only supported by the major FPGA
vendors, but also a major technology used for the programming of GPUs,
with the corresponding programming environments and experience in
the community. In this paper, we, therefore, employ OpenCL for the
development of relatively simple signal processing tasks on FPGAs.
We want to explore the advantage of a high-level approach to cover
a large design space, by testing many different approaches. It will
be interesting to see whether OpenCL can exploit the FPGA resources
efficiently for this task. We are encouraged by the successful use
of the high-level development technique in many research areas regarding
hardware acceleration, such as high-speed data compression~\cite{abdelfattah2014gzip},
Map/Reduce, and computationally demanding control
algorithms~\cite{navarro2013high}.

\section{FDAS Module and FT Convolution}\label{FDAS-Module}

\begin{figure}
\begin{centering}
\includegraphics[bb=10bp 10bp 550bp 180bp,clip,scale=0.55]{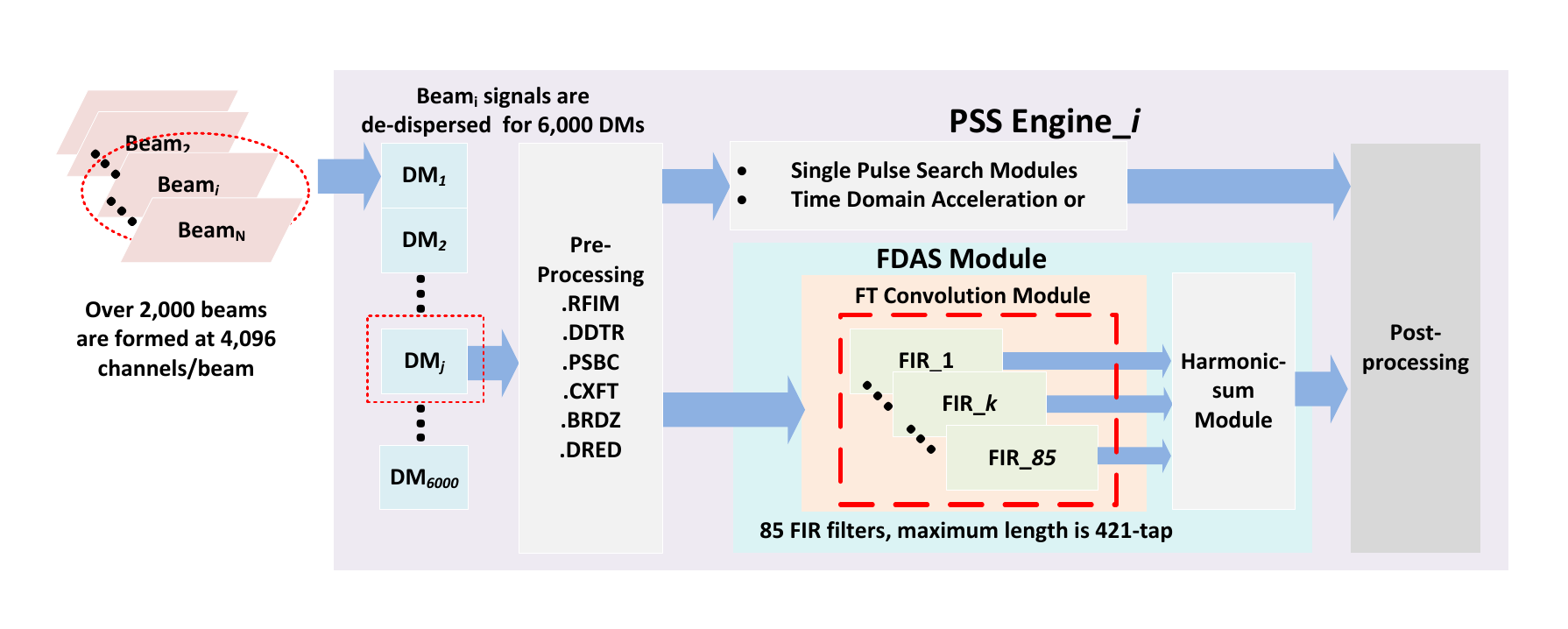}
\par\end{centering}

\protect\caption{\label{fig:Data-flow-of}Data flow of SKA1-MID CSP PSS engine}
\end{figure}

In the SKA1-MID CSP element, over 2,000 beams are formed at 4,096
channels per beam, and the signals of each beam are processed independently,
as depicted in Figure~\ref{fig:Data-flow-of}. Hence, each beam needs
a dedicated pulsar search engine (PSS). Because the dispersion measure
(DM) is unknown (we are looking for pulsars at unknown locations),
about 6,000 trial values are tested, and several pulsar search approaches
are employed for each trial value. These approaches include single
pulse search module, time domain acceleration search module, and frequency
domain acceleration search (FDAS) that we are investigating. 

The FDAS module consists of two main sub-modules: the FT convolution
module and the harmonic summing module. In the FT convolution module,
85 templates with different lengths are applied in each trial. The
$2^{22}$ input points, which data type is complex single precision
floating-point (SPF), are accumulated in an integration time of $536.87s$.
With 6,000 DM trials to perform until the next input set is ready,
the time limitation $t_{limit}$ for processing each DM trial is $89.5ms$
(536.87/6,000). In this $89.5ms$, the main computing task is to convolve
the $2^{22}$ complex SPF points with 84 templates and calculate the
power of each complex output points. Each template can be seen as
an FIR filter, whose input signals, coefficients, and output points
are all complex SPF points. The output points from all 84 FIR filters
plus the input signals are combined into a filter-output-plane (FOP,
85$\times$4-million complex points) and the spectral\textbf{ }power
of each complex point is sent to the harmonic-summing module for candidate
detection. The process that described above is illustrated Figure~\ref{fig:Data-flow-of}.
In these 84 FIR filters, the lengths of them are different, and the
longest FIR filter has 421 taps. In case of the uncertainty of the
FIR filter length, we investigate the implementation of 84 421-tap
FIR filters in this research. 

The real-time compute-intensive task of the FT convolution module
described above is a large challenge for efficient computation, which
essentially makes the use of acceleration hardware necessary. Based
on the specifications of the FT convolution module, the performance
needed of the 84 FIR filters is
\[
\frac{8NKM}{t_{limit}}=13.26TFLOPS,
\]
 where $N=2^{22}$ is the input size, $K=421$ is the length of each
FIR filter, $M=84$ is the number of FIR filters, and eight is derived
that one complex multiplication needs eight operations (four multiplications
and four additions).The $13.26TFLOPS$ per beam is based on the straightforward
implementation of 1D convolution. Since there are over 2,000 beams,
the overall needed performance for one pulsar search module is over
$26.5PFLOPS$. In \cite{wang2015fpga}, relaxation of requirements
was studied to ease the requirements, such as changes to the input
data type, size, number of filters, etc. From this early work, it
was clear that very efficient implementations of the filtering task
need to be investigated. In this section and the remainder of this
paper, we, therefore, investigate different algorithms for FPGA-based
acceleration of the FT convolution of the FDAS module. Based on the
processing domain of an FIR filter, this is divided into time-domain
and frequency-domain.

\subsection{Time-domain FIR Filter (TDFIR)}\label{subsec:Time-domain-FIR-Filter}

\subsubsection{Na{\"i}ve TDFIR}

Based on the discrete-time convolution, an $K-$tap FIR filter can
be represented as 
\begin{equation}
y[i]=\sum_{k=0}^{K-1}x[i-k]h[k],\,for\,i=0,1,\,...N-1,\label{eq:tdfir}
\end{equation}
 where $x[\cdot]$, $h[\cdot]$, and $y[\cdot]$ are complex SPF input
signals, coefficients, and output results, respectively, and $N$
is the input size~\cite{smith1997scientist}. In an FPGA implementation,
the SPF multipliers are instantiated by DSP blocks and logic resources.
If there are enough resources on an FPGA, the $K$ multiplications
and additions in (\ref{eq:tdfir}) can be parallelised in a pipeline
completely to achieve high-performance.

\subsubsection{Overlap-add Algorithm based TDFIR}

The amount of logic resource and DSP blocks in a specific FPGA are
fixed. If the FIR filter size $K$ is too large, an FPGA might not
have enough logic resources and DSP blocks to parallelise $K$ complex
multiplications and then fails to achieve a pipeline structure. To
make an FIR filter fit into the targeted FPGA and maintain high-performance,
we apply the overlap-add algorithm (\textbf{OLA}) to split the coefficient
array into a group of sub-arrays~\cite{pavel2013algorithms}.

\begin{algorithm}
\protect\caption{\label{alg:Overlap-add-Algorithm}Overlap-add Algorithm}

{\small{}$(x,\,h)\leftarrow$(padded input data, coefficients)}{\small \par}

{\small{}$h\rightarrow h_{1},\,h_{2},\,...,\,h_{R}$ \{split $h$
into a group of $R$ disjoint sub-arrays evenly\}}{\small \par}

{\small{}$y\leftarrow0$ \{create the output $y$ and fill it with
zeros\}}{\small \par}

\textbf{\small{}for $i=1$ }{\small{}to $R$ }\textbf{\small{}do}{\small \par}

{\small{}~~~~$y_{i}\leftarrow$convolve$(x,\,h_{i})$ \{general
convolution in time-domain\}}{\small \par}

{\small{}~~~~$y\leftarrow y+$shift$(y_{i})$ \{add $y_{i}$ to
$y$\}}{\small \par}

\textbf{\small{}end for}{\small \par}

\end{algorithm}

Algorithm~\ref{alg:Overlap-add-Algorithm} and Figure~\ref{fig:The-Process-of}(a)
outline the OLA process of splitting an FIR filter into $R$ small
sub-FIR filters. The coefficient array is evenly decomposed into $R$
disjoint sub-arrays. If filter size $K$ as $K=R\times K'$, where
$R$ and $K'$ are integers, $K'-1$ zero points will be padded at
the end of the input array in the OLA algorithm. Each output array
has to be shifted by $K'$ points and then added to the previous output
array.

\subsection{\label{subsec:Frequency-domain-FIR-Filter}Frequency-domain FIR Filter
(FDFIR)}

\subsubsection{\label{subsec:Na=0000EFve-FDFIR}Na\text{\"i}ve FDFIR }

Based on the convolution theorem, Equation (\ref{eq:fdfir}), the
output of an FIR filter can be obtained by the following steps~\cite{smith1997scientist}:
Fourier transform of the input array and coefficient array, element-wise
multiplication of these two arrays, and inverse Fourier transform
of the output array. 
\begin{equation}
x\ast h=\mathcal{F}^{-1}\{\mathcal{F}\{x\}\cdot\mathcal{F}\{h\}\},\label{eq:fdfir}
\end{equation}
where $\mathcal{F}\{\cdot\}$ and $\mathcal{F}^{-1}\{\cdot\}$ are
Fourier transform and inverse Fourier transform. For FPGA implementations,
a fast Fourier transform (FFT) engine will be instantiated to handle
Fourier and inverse Fourier transform.

\subsubsection{Overlap-save Algorithm based FDFIR}

\begin{figure}
 \begin{centering}
\includegraphics[bb=0bp 40bp 810bp 300bp,clip,scale=0.5]{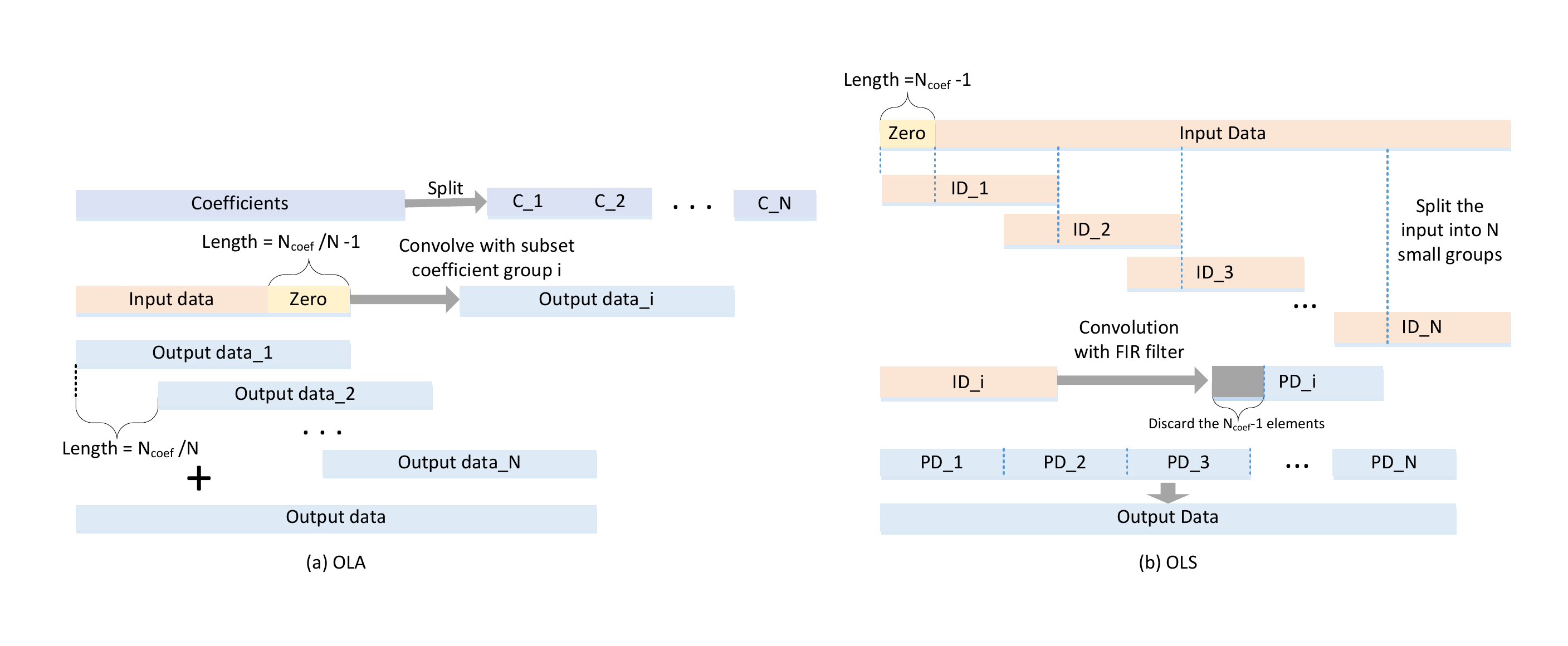}
 \par\end{centering}

 \protect\caption{\label{fig:The-Process-of}Process of the OLA and OLS algorithms}
 \end{figure}

For Fourier transforming large size input, such as the targeted four
million points ($2^{22}$) FFT, the on-chip memory of an FPGA is unable
to store all points, which makes it impossible to perform the complete
process as described in Section~\ref{subsec:Na=0000EFve-FDFIR} in
one go. Hence, we apply the overlap-save algorithm (OLS) to split
the input signals into chunks~\cite{pavel2013algorithms}. Each chunk
overlaps with its two neighbour chunks, and the extent of the overlap
is $K-1$, where $K$ is the FIR filter length. For the first input
chunk, $K-1$ zero points have to be padded at the beginning. After
convolving in frequency-domain, the overlap, which is the first $K-1$
points of each chunk, are discarded. The OLS algorithm in Algorithm~\ref{alg:Overlap-save-Algorithm}
and Figure \ref{fig:The-Process-of}(b) illustrate the process of
splitting the input array of size $N$ into $S$ sub-arrays~\cite{pavel2013algorithms}
of size $N/S$.

\begin{algorithm}
\protect\caption{\label{alg:Overlap-save-Algorithm}Overlap-save Algorithm}

{\small{}$(x,\,h)\leftarrow$(padded input data, coefficients)}{\small \par}

{\small{}$x\rightarrow x_{1},\,x_{2},\,...,\,x_{S}$ \{split $x$
into group of sub-arrays; successive sub-arrays overlap by $K-1$
points\}}{\small \par}

{\small{}$y\leftarrow0$ \{initialize output $y$ with zeros\}}{\small \par}

{\small{}$\mathscr{F}\{h\}\leftarrow$FFT$(h)$ \{Fourier transform
of coefficient array\}}{\small \par}

\textbf{\small{}for $i=1$ }{\small{}to $S$ }\textbf{\small{}do}{\small \par}

{\small{}~~~~$\mathscr{F}\{x_{i}\}\leftarrow$FFT$(x_{i})$ \{Fourier
transform of input sub-array $x_{i}$\} }{\small \par}

{\small{}~~~~$\mathscr{F}\{y_{i}\}\leftarrow$times($\mathscr{F}\{x_{i}\},\,\mathscr{F}\{h\}$)
\{element-wise multiplication of Fourier transformed input sub-array
and coefficient array\}}{\small \par}

{\small{}~~~~$y_{i}\leftarrow$IFFT$(\mathscr{F}\{y_{i}\})$ \{inverse
Fourier transform\}}{\small \par}

{\small{}~~~~$y_{i}\leftarrow$discard\_overlap ($y_{i}$) \{discard
the front $K-1$ points\}}{\small \par}

{\small{}~~~~$y\leftarrow$shift$(y_{i})$ \{concatenate $y_{i}$
to $y$\}}{\small \par}

\textbf{\small{}end for}{\small \par}

{\small{}Output$\leftarrow y$}{\small \par}
\end{algorithm}

\subsection{Workload Comparison}

The main advantage of FDFIR is that its workload growth is slower
than that of TDFIR as the FIR filter length increases. Assuming that
the input length $N$ is several magnitudes larger than the FIR filter
length $K$, which means $N\gg K$ and $N+K-1\approx N$. This holds
for our FDAS filtering task. Table~\ref{tab:Workload-of-one} compares
the workload of the four different approaches discussed in the previous
sections. In the FDFIR, the complexity of computing the FFT is $O(Nlog_{2}N)$.
For single complex multiplication, six general operations are needed
(four multiplications and two additions), and two additional additions
for the accumulation are used for summing in case of TDFIR. 

\begin{table}
\tbl{Workload of TDFIR and FDFIR\label{tab:Workload-of-one}}{%

\centering{}%
\begin{tabular}{c|c|cc}
\hline 
{\scriptsize{}Domain} & {\scriptsize{}Algorithm} & {\scriptsize{}Workload (Single filter)} & {\scriptsize{}Average Workload ($M$ filters)}\tabularnewline
\hline 
\multirow{2}{*}{{\scriptsize{}TD}} & {\scriptsize{}Naive} & {\scriptsize{}$8KN$ } & {\scriptsize{}$8KN$}\tabularnewline
 & {\scriptsize{}OLA} & {\scriptsize{}$8K'N\left\lceil \frac{K}{K'}\right\rceil $} & {\scriptsize{}$8K'N\left\lceil \frac{K}{K'}\right\rceil $}\tabularnewline
\multirow{2}{*}{{\scriptsize{}FD}} & {\scriptsize{}Naive} & {\scriptsize{}$N(6+2C\cdot log_{2}N)$} & {\scriptsize{}$N(6+C\cdot log_{2}N)$}\tabularnewline
 & {\scriptsize{}OLS} & {\scriptsize{}$\left\lceil \frac{N}{N_{FT}-K}\right\rceil N_{FT}(6+2C\cdot log_{2}N_{FT})$} & {\scriptsize{}$\left\lceil \frac{N}{N_{FT}-K}\right\rceil N_{FT}(6+C\cdot log_{2}N_{FT})$}\tabularnewline
\hline 
\end{tabular}}
\end{table}

In Table~\ref{tab:Workload-of-one}, $K'$ denotes the tap length
that an FPGA can parallelise completely (also the sub-FIR filter length
in Algorithm~\ref{alg:Overlap-add-Algorithm}), $N_{FT}$ is the
Fourier transform length (also the chunk length in Algorithm~\ref{alg:Overlap-save-Algorithm}),
and $C$ is a constant depending on the applied FFT algorithm, which
is typically less than 5. For the FD algorithms, the one-off workload
cost for the Fourier transform of the coefficient array is not included,
as it is negligible with the assumption $N\gg K$. The table compares
the workload for a single filter and the average workload for $M$
filters ($M\gg1$), which is relevant to the FT convolution module.
The essential difference is that the forward Fourier transform only
needs to be performed once for all $M$ filters.

For the OLA-TD, if $K$ can be divided by $K'$, the workload equals
to the Na\text{\"i}ve-TD workload. For FD algorithms, the workload
of Na\text{\"i}ve-FD is not affected by FIR filter length $K$, but
the workload of the OLS-FD will rise with the filter size $K$. When
$K$ is fixed, the smaller the $N_{FT}$, the larger the overall workload.
However, if the $N_{FT}$ is too large, the performance might drop
because of the on-chip memory size and the off-chip access efficiency.
Hence finding the suitable $N_{FT}$ for a specific FPGA device is
investigated in the evaluation section. 

Based on the theoretical workload, FD algorithms have a clear advantage
over TD algorithms in implementing multiple FIR filters. The evaluation
will show if this advantage can be achieved in practice. 

\section{FPGA-based FIR Filter Structure}\label{FIR_Structure}

In this section, we discuss the structures of the two non-na\text{\"i}ve
FIR filters introduced in Sections~\ref{subsec:Time-domain-FIR-Filter}
and \ref{subsec:Frequency-domain-FIR-Filter}: the OLA algorithm based
TDFIR, referred to as \textbf{OLA-TD}, and the OLS algorithm based
FDFIR, referred to as \textbf{OLS-FD}. 

\subsection{\label{subsec:OLA-TD}OLA-TD}

To handle large size FIR filters, we investigate here the structure
of OLA-TD. It is based on the Na\text{\"i}ve TDFIR, the input signals
are loaded into a shift register, the core computation part is entirely
parallelised using DSP blocks, and the structure can achieve loop
pipelining. However, each output array of the OLA-TD is shifted and
accumulated to the previous output array. 

Assuming an FPGA can completely parallelise $K^{'}$ complex SPF multiplications,
then the same input array has to be executed by OLA-TD structure $R$
times to implement a $K$-tap FIR filter, where $R=\left\lceil \frac{K}{K'}\right\rceil $.
For the Intel Stratix V FPGAs, one complex SPF multiplication needs
four DSP blocks, so $K^{'}$ is decided by the number of DSP blocks
on an FPGA. If there are $N_{DSP}$ DSP blocks, then $K^{'}=\left\lfloor \frac{N_{DSP}}{4}\right\rfloor $.
Because of loop pipeline, it takes $R\times N$ clock cycles to process
$N$ points with a $K$-tap FIR filter.

\subsection{\label{subsec:OLS-FD}OLS-FD}

For the OLS-FD structure, two important components are Fourier transform
and inverse Fourier transform. In our work, we employ a complex SPF
radix-4 feedforward FFT/IFFT engine~\cite{garrido2013pipelined}
provided by Intel. The single FFT engine can be configured so that
it provides both FFT and IFFT. The input signals are in general order,
and the output array is in bit-reversed order. The workload (number
of operations) of the employed FFT engine in processing $N_{FT}$
points is $5N_{FT}log_{2}N_{FT}$, so here the constant $C$ in Table~\ref{tab:Workload-of-one}
is 5. It can process multiple points, referred to as $N_{FT-PC}$,
per clock cycle, where $N_{FT-PC}$ is a power of 2 such as 4 and
8. It takes the engine $\frac{N_{FT}}{N_{FT-PC}}-1$ clock cycles
to produce $N_{FT-PC}$ points output for a corresponding $N_{FT-PC}$
points input. If the output array has to be of general order, a bit-reverse
module needs to be added after the FFT engine to reorder the output
array. 

There are two factors that limit the FFT engine on a specific device:
the bandwidth of off-chip memory and the number of DSP blocks. The
bandwidth of off-chip memory limits the number of points that can
be loaded and stored per clock cycle, and the amount of DSP blocks
determines the number of instantiated FFT engines. Two different OLS-FD
structures are proposed differing in the number of instantiated FFT
engines: area-efficient OLS-FD (\textbf{AOLS-FD}) and time-efficient
OLS-FD (\textbf{TOLS-FD}).

\subsubsection{AOLS-FD}

The AOLS-FD structure, as shown in Figure~\ref{fig:Structure-of-time-efficient}(a),
consists of three separate parts: data fetch and multiplication, FFT/IFFT,
and bit-reverse. These three parts are connected through FIFO buffers
in an FPGA. The core computation part of it is the reconfigurable
FFT engine. The multiplication performed in data fetch and multiplication
part is the element-wise multiplication. To process one input array,
the AOLS-FD structure has to be executed twice. For the first time,
the input array and initial array are stored in one off-chip memory
bank (Bank$_{1}$) and the Fourier transformed coefficient array is
stored in another memory bank (Bank$_{2}$). The initial array is
an array that is initialized with neutral elements ($1+j\cdot0$)
so that the multiplication does not affect in the first round. After
the first execution, the intermediate data generated by the bit-reverse
kernel are stored in Bank$_{2}$. In the second round, the FFT engine
is configured as IFFT. The data fetch and multiplication part loads
intermediate output array and pre-processed coefficient array from
Bank$_{2}$ and the bit-reversed kernel stores the final output array
in Bank$_{1}$. The first round is only necessary once for multiple
filters, hence becomes less important with growing $M$, which are
discussed in Section~\ref{subsec:Multiple-FIR-Filters}.

\subsubsection{TOLS-FD}

\begin{figure}
\begin{centering}
\includegraphics[bb=0bp 15bp 474bp 380bp,clip,scale=0.7]{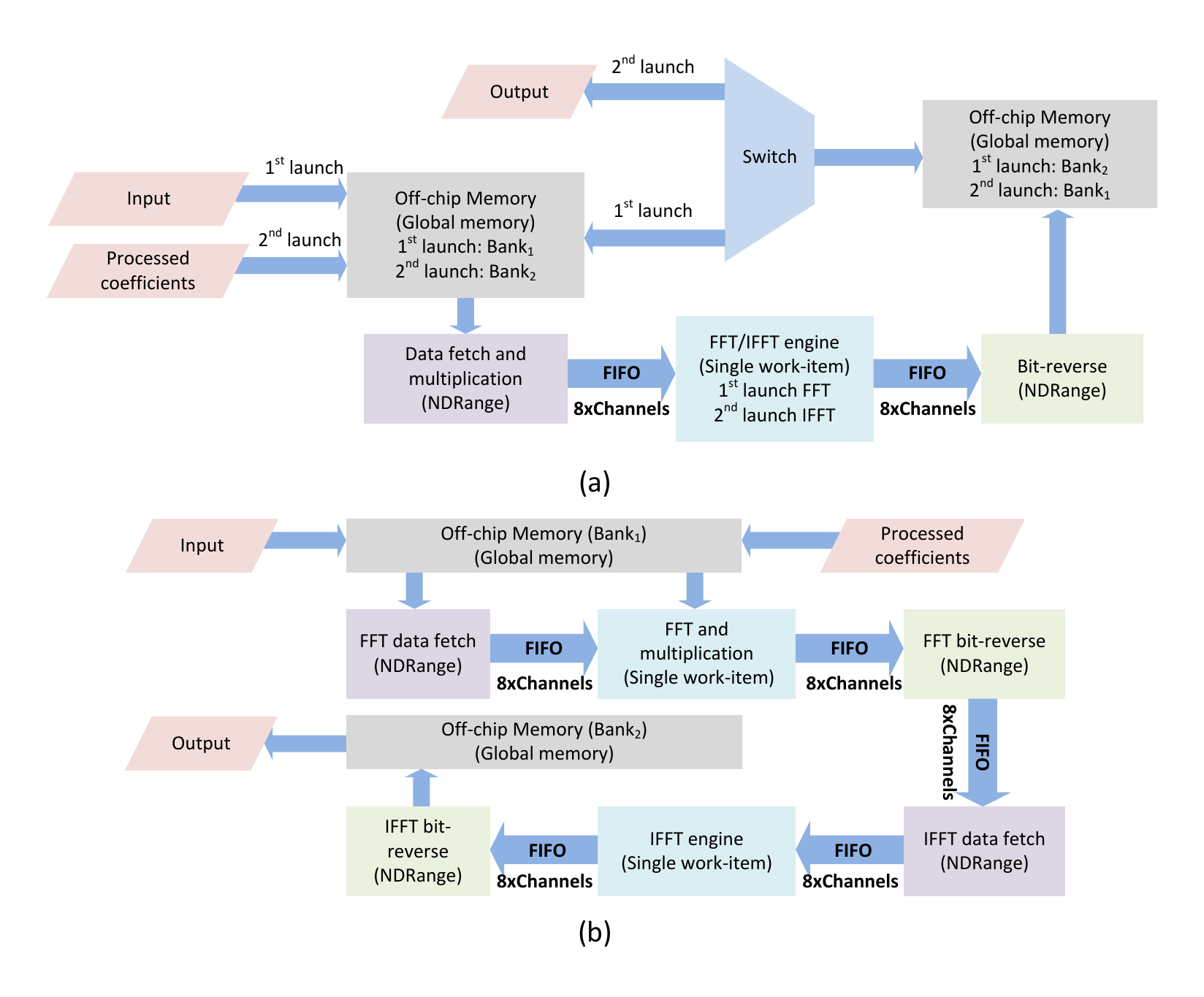}
\par\end{centering}
\caption{\label{fig:Structure-of-time-efficient}Structures of (a) AOLS-FD
and (b) TOLS-FD}
\end{figure}

The TOLS-FD structure, Figure~\ref{fig:Structure-of-time-efficient}(b),
is based on the AOLS-FD structure, however, it only needs to be executed
once. TOLS-FD contains two FFT engines, one for FFT and another for
IFFT. In this case, there is no need to store the intermediate results
in off-chip memory, which reduces the frequency of memory usage. Different
from the AOLS-FD structure, the element-wise multiplication is put
after FFT engine and before bit-reverse, so the pre-processed Fourier
transformed coefficient array needs to be in bit-reversed order. The
input array and pre-processed coefficient array are stored in Bank$_{1}$
and the output array is stored in Bank$_{2}$. 

\subsection{\label{subsec:Optimisation-for-FT}Optimisation for FT Convolution
Module}

The resource usage of the acceleration device plays an important role
when multiple FIR filters need to be processed as it is meaningful
in parallelising the design. The off-chip memory bandwidth and the
amount of DSP blocks $N_{DSP}$ are two factors that affect the proposed
structures.

In this paper, we employ the Terasic DE5-Net acceleration card as
the target acceleration device, and the detailed analysis is based
on it. A DE5 card has one Intel Stratix V 5SGXA7 FPGA, referred to
as $\mathbf{A7}$. The $A7$ FPGA possesses 256 DSP blocks, which
means 256 SPF or 64 complex SPF multiplications can be performed in
parallel. The DE5 card has two DDR3 memory banks (Bank$_{1}$ and
Bank$_{2}$) and each bank is connected with $A7$ FPGA through a
64-bit data bus. The maximum frequency of DDR3 SDRAM supported by
the DE5 card is $1,066MHz$, thus the theoretical peak data transfer
rate for one memory bank is $64\times2\times800=136Gbps$ (2 for double
data rate).

For the TDFIR, $\frac{N_{DSP}}{4}$, which is 64, is smaller than
the FIR filter length $K$, which is 421, so only one input signal
can be processed per clock cycle using the OLA-TD method\@. During
processing, one input signal is loaded, and one result point is stored
in one clock cycle, which needs 128 bits/cycle in total. Based on
the theoretical transfer rate, the off-chip memory will become the
barrier only when the FPGA operation frequency is higher than $1GHz$,
which is impossible for Stratix V FPGAs.

\begin{figure}
\begin{centering}
\includegraphics[bb=10bp 10bp 540bp 350bp,clip,scale=0.65]{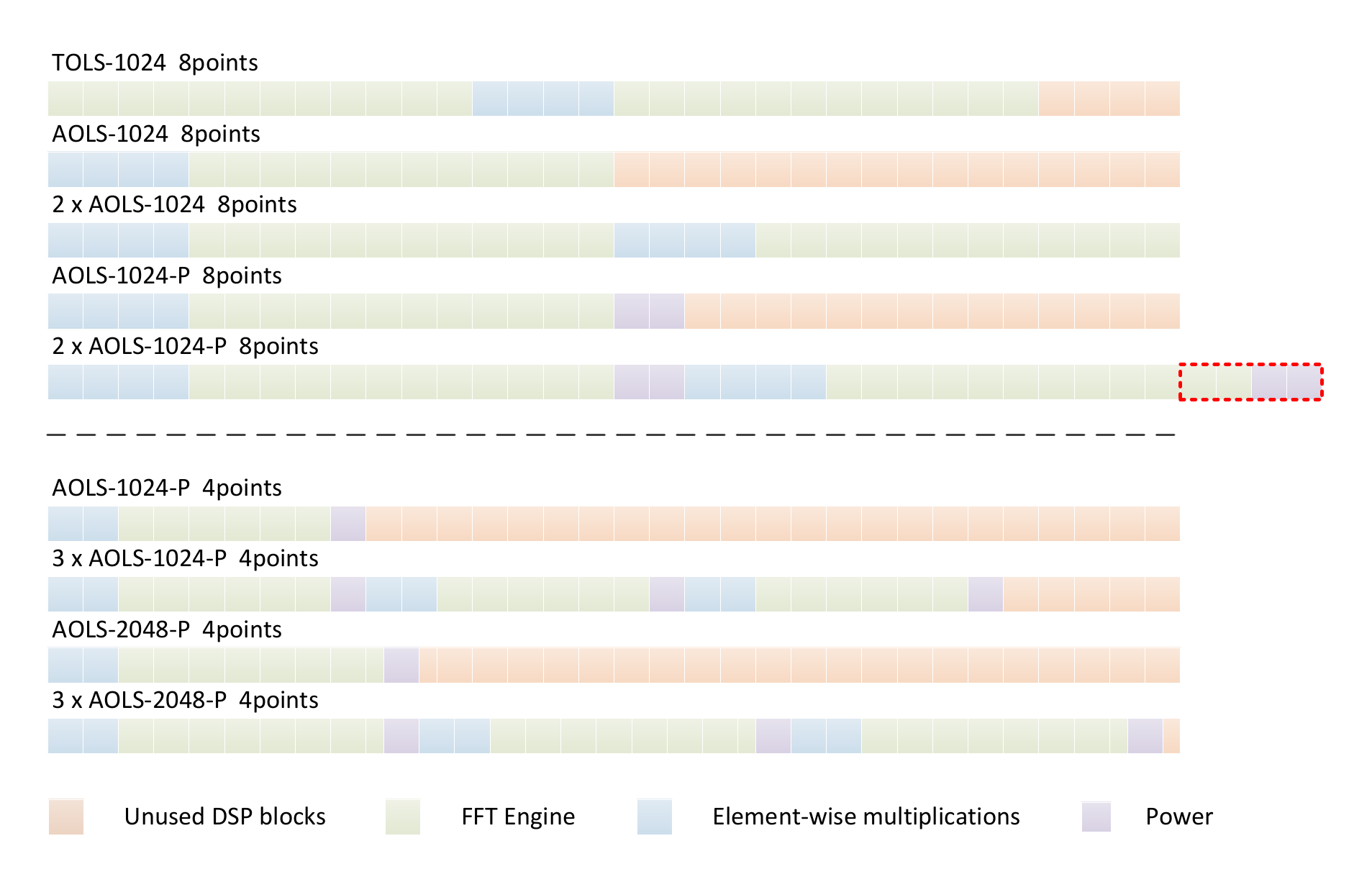}
\par\end{centering}
\caption{\label{fig:DSP-blocks-usage}DSP blocks usage of OLS-FD structures}
\end{figure}

Regarding the FDFIR, the segment length $N_{FT}$ and $N_{FT-PC}$
are two important parameters that influence both TOLS-FD and AOLS-FD
structures. Let us look at the DSP block usage of TOLS-$N_{FT}$ and
AOLS-$N_{FT}$ using 8 points FFT engine on an $A7$ FPGA as depicted
in Figure~\ref{fig:DSP-blocks-usage}. The DSP cost of the FFT engine
is decided by $N_{FT}$ and $N_{FT-PC}$, and the cost of element-wise
multiplication part is decided by $N_{FT-PC}$ only. We see a symbolic
representation of the DSP block consumption of the different settings,
distinguished by the components of the structures as FFT engine, element-wise
multiplication, etc.

For an 8-point 1,024 FFT engine, it has 96 multiplications that cost
96 DSP blocks. In Figure~\ref{fig:DSP-blocks-usage}, TOLS-1024 consumes
224 ($96\times2+32$, which costs 88\% of overall DSP blocks) DSP
blocks and such an implementation takes a large amount of off-chip
memory bandwidth, which is 1,024 bits/cycle (8 points$\times$64-bit$\times$2).
AOLS-1024 only consumes 128 (50\%) DSP blocks, and therefore it is
possible to parallelise two AOLS-1024 structures on one $A7$ FPGA.
However, this increases the required off-chip memory bandwidth to
1,536 bits/cycle (8 points$\times$64-bit$\times$3). Since the theoretical
peak data transfer is fixed, the increase of the required off-chip
memory bandwidth leads to the decrease of FPGA operation frequency.
Due to that, the performance of two AOLS-1024 (2 x AOLS-1024 using
8 points FFT engine) might not be 2x times faster than that of a single
AOLS-1024 on an $A7$ FPGA. 

In the FT convolution module, only the spectral power values of
the FIR filter (i.e., one SPF) are required, which essentially halves
the output bandwidth requirement. Calculating the power of each complex
value requires simple floating-point multiplications (the square root
is not necessary for this processing) which consume some more DSPs.

We name \textbf{AOLS-$N_{FT}$-P} to represent the AOLS-$N_{FT}$
based structure while calculating the power of complex value. By calculating
the power, the required bandwidth of 8-point FFT engine based structure
is reduced from $1,024$ bits/cycle to $768$ bits/cycle. Although
it is possible for an $A7$ FPGA to parallelise two AOLS-1024 structures,
there are no more DSP blocks to calculate the power of complex value.
This is indicated by the overrun on the red dot part in Figure~\ref{fig:DSP-blocks-usage}.
The numbers of DSP blocks used for implementing element-wise multiplication
and power calculation are decided by the number of processed points
per clock cycle $N_{FT-PC}$ of the FFT engine. 

To best exploit the resources on the FPGA, we reduce the points that
are simultaneously processed by the FFT engine $N_{FT-PC}$ from eight
to four, and the resources usage of such kernels is illustrated in
the bottom half of Figure~\ref{fig:DSP-blocks-usage}. It can be
seen that up to three AOLS-1024-P or AOLS-2048-P structures can be
instantiated in parallel on an $A7$ FPGA. By employing the 4-point
FFT engine, the simultaneously processed points increased from 8 (a
single 8-point FFT engine) to 12 (3$\times$4-point FFT engines).
The required off-chip memory bandwidth for 4-point FFT engine based
3x AOLS-$N_{FT}$-P structure is reduced from 1,024 bits/cycle (AOLS-$N_{FT}$
using 8-point FFT engine and without the power computation) to 640
bits/cycle (4 points$\times$64-bit$+$4 points$\times$32-bit$\times3$).

\section{High-level Approaches and Implementation}\label{Architecture-and-Optimisation}

For the implementation of the various FIR filtering algorithms, we
have selected the high-level approach of using OpenCL. It is employed
for multiple purposes, which are as follows:
\begin{itemize}
\item It allows for fast prototyping of the proposed implementations and
to systematically explore a large design space while achieving high-performance
computing;
\item It provides portability to other platforms (i.e., CPUs and GPUs) and
between generations of the same devices (e.g., Stratix V and Arria
10);
\item It makes the developed implementation accessible to non-hardware-design
experts which is essential in such a large cross-discipline project
as the SKA.
\end{itemize}

\subsection{OpenCL for FPGA}

The open computing language (OpenCL) is based on standard ANSI C (C99)
and can be executed on heterogeneous platforms. The OpenCL platform
consists of two components: host and devices~\cite{khronosopencl}.
In our research, one FPGA board can be seen as a compute device, and
multiple acceleration cards can be installed to the host processor.
OpenCL uses the concept of two types of memory, local (fast) memory
and global (slower) memory. The off-chip memory of the FPGA board
such as SDRAM and QDRII SRAM is instantiated as the global memory
of OpenCL kernels and FPGA on-chip memory such as BRAM is used as
the local memory of OpenCL kernels. 

Developing for FPGA(s) using OpenCL mainly contains two parts: OpenCL
kernels for the devices and software programs for the host. To compile
the FPGA-targeted OpenCL kernels, a dedicated compiler is required.
In our research, the Altera SDK for OpenCL (AOCL) is employed, and
the AOCL offline compiler (AOC) is used. The current AOCL conforms
to OpenCL specification version 1.0 and some functions of versions
1.2 and 2.0, such as \texttt{clGetKernelArgInfo} and pipe~\cite{altera2016openclpro}.
Furthermore, the AOCL has optimization techniques for FPGAs, including
unrolling \texttt{for} loops, using channels to connect different
kernels, and optimizing floating-point operations. The FPGA-based
OpenCL kernels can be distinguished into two types: single work-item
kernels, which are recommended by Altera~\cite{altera2016openclpra},
and NDRange kernel, where several work-items are processed together. 

The host programs, which is written in C or C++, are responsible for
the management and the remaining work, such as organizing data between
the host processor and device(s), setting the arguments of OpenCL
kernels, and launching kernels. In practical execution, before launching
OpenCL kernels, related data arrays are transferred from the host
processor to the global memory of the FPGA board through the PCI Express
(PCIe) bus~\cite{czajkowski2012opencl}. Multiple devices can be
connected to the host through the PCIe bus. When an OpenCL kernel
is launched, it loads data into the global memory of the FPGA device.
Depending on the kernel\textquoteright s function, part of the data
or intermediate results might be stored in local memory. After executing,
the output array that is stored in the global memory is sent back
to the host processor.


\subsection{FIR Filter Kernel}

Based on the discussed structures in Section~\ref{FIR_Structure},
we investigate the implementations of these using OpenCL kernels.

\subsubsection{OLA-TD\label{sub:OLA-TD}}

The OLA-TD kernel can be implemented as both the single work-item
kernel or NDRange kernel, and the difference between these two kernel
types are investigated in this paper. The kernel codes are given in
Figure~\ref{fig:OpenCL-Code-(using}, where we set $K'=64$ (SFL).
The core computation part can be completely unrolled by adding \texttt{\#pragma
unroll}. For the NDRange kernel, the global work size that is defined
in the host program is the length of input signals, and the work-group
size is specified in the device kernel (\texttt{\_\_attriute\_\_((reqd\_work\_group\_size(SFL,~1,~1)))}),
i.e., here 64 (SFL) in one dimension.

Although the NDRange kernel executes the same amount of complex SPF
multiplication per clock cycle as the single work-item kernel, their
structures are different. For the NDRange OLA-TD kernel, 64 work-items
compose one work-group. By using the OpenCL barrier (\texttt{barrier()}),
all the related input of one work-group have to be loaded before executing
the core computation part. The FPGA executes all work-groups sequentially,
and for each work-group, one work item will be executed every clock
cycle.

\begin{figure}[htbp]
\begin{minipage}[t]{0.48\linewidth}
\begin{lstlisting}
#define SFL 64  //Sub-Fliter Length

__attribute__((task)) 
__kernel void OLA-64S (         
__global float *restrict dataPtr,  //Input
__global float *restrict filterPtr,//Coefficients
__global float *restrict resultPtr,//Output		         
const int offset, // 1 to num_chunks
const int num_chunks,       
const int totalInputLength){   

float i_re[SFL];   
float i_im[SFL];   
int ilen, k;    
    
#pragma unroll       
  for (k=0; k < SFL; k++){         
    coef_real[k] = filterPtr[2*(SFL*(offset-1)+k)]; 
    coef_imag[k] = filterPtr[2*(SFL*(offset-1)+k)+1]; 
  }

// Main loop to process input signals  
  for(ilen = 0; ilen < totalInputLength; ilen++){     
    float r_re = resultPtr[2*(ilen + offset * SFL)];     
    float r_im = resultPtr[2*(ilen + offset * SFL)+1];    
#pragma unroll     
    for (k=0; k < SFL-1; k++){       
      i_re[k] = i_re[k+1];       
      i_im[k] = i_im[k+1];     
    }
    // Shift in 1 complex data point to process     
    i_re[SFL-1] = dataPtr[2*ilen];      
    i_im[SFL-1] = dataPtr[2*ilen+1];      
    
    //unroll core computation part of OLA-TD     
#pragma unroll     
    for (k=SFL-1; k >=0; k--){       
      r_re += i_re[k] * coef_real[SFL-1-k]  
              - i_im[k] * coef_imag[SFL-1-k];       
      r_im += i_re[k] * coef_imag[SFL-1-k]  
              + i_im[k] * coef_real[SFL-1-k];     
    }     
    resultPtr[2*(ilen + offset*SFL)] = r_re;     
    resultPtr[2*(ilen + offset * SFL)+1] = r_im;   
  } 
}
\end{lstlisting}
\end{minipage}
\hfill
\begin{minipage}[t]{0.48\linewidth}
\begin{lstlisting}
#define SFL 64  //Sub-Fliter Length

__attribute__((reqd_work_group_size(SFL, 1, 1))) 
__kernel void OLA-64N(
__global float *restrict dataPtr,  //Input
__global float *restrict filterPtr,//Coefficients
__global float *restrict resultPtr,//Output
const int offset, // 1 to num_chunks
const int num_chunks){

unsigned int i_g = get_global_id(0);
unsigned int i_l = get_local_id(0);  
unsigned int load_i = i_g + (offset - 1)*SFL;

//Load input to local memory (on-chip)
__local float i_re[SFL*2];    
__local float i_im[SFL*2];
i_re[i_l] = dataPtr[2*load_i];
i_im[i_l] = dataPtr[2*load_i + 1];
i_re[i_l+SFL] = dataPtr[2*(load_i+SFL)];
i_im[i_l+SFL] = dataPtr[2*(load_i+SFL) + 1];

//To make sure the local memory is loaded properly
barrier(CLK_LOCAL_MEM_FENCE);


//Load the results generated during last launch
float r_re = resultPtr[2*i_g];
float r_im = resultPtr[2*i_g + 1];
unsigned int tap = (num_chunks - offset)*SFL - 1;


#pragma unroll   
for(unsigned int ilen = 0; ilen<SFL; ilen++){
load_i = i_l + ilen;
  r_re += i_re[load_i]*filterPtr[2*(tap-ilen)]
          - i_im[load_i]*filterPtr[2*(tap-ilen)+1];
  r_im += i_re[load_i]*filterPtr[2*(tap-ilen)+1]
          + i_im[load_i]*filterPtr[2*(tap-ilen)];	
}


//Save the output
resultPtr[2*i_g] = r_re;   
resultPtr[2*i_g + 1] = r_im;
}; 


\end{lstlisting}
\end{minipage}

\protect\caption{\label{fig:OpenCL-Code-(using}OpenCL code (using Single work-item(Left) and NDRange(Right)) of OLA-TD kernel}
\end{figure}

\subsubsection{OLS-FD}

Based on the structures of AOLS-FD and TOLS-FD in Figure~\ref{fig:Structure-of-time-efficient},
they can be implemented using OpenCL kernels directly. Each function
block is a kernel, and different function kernels are connected with
the channels. The data fetch and multiplication kernel and bit-reverse
kernel are simple that can be implemented using NDRange kernel. For
the FFT engine, the single work-item kernel type is employed. To execute
the same amount of input signals, AOLS-$N_{FT}$ kernel needs to be
launched twice, and TOLS-$N_{FT}$ kernel only needs to be launched
once. 

\begin{figure}
\begin{centering}
\includegraphics[bb=0bp 30bp 595bp 390bp,clip,scale=0.65]{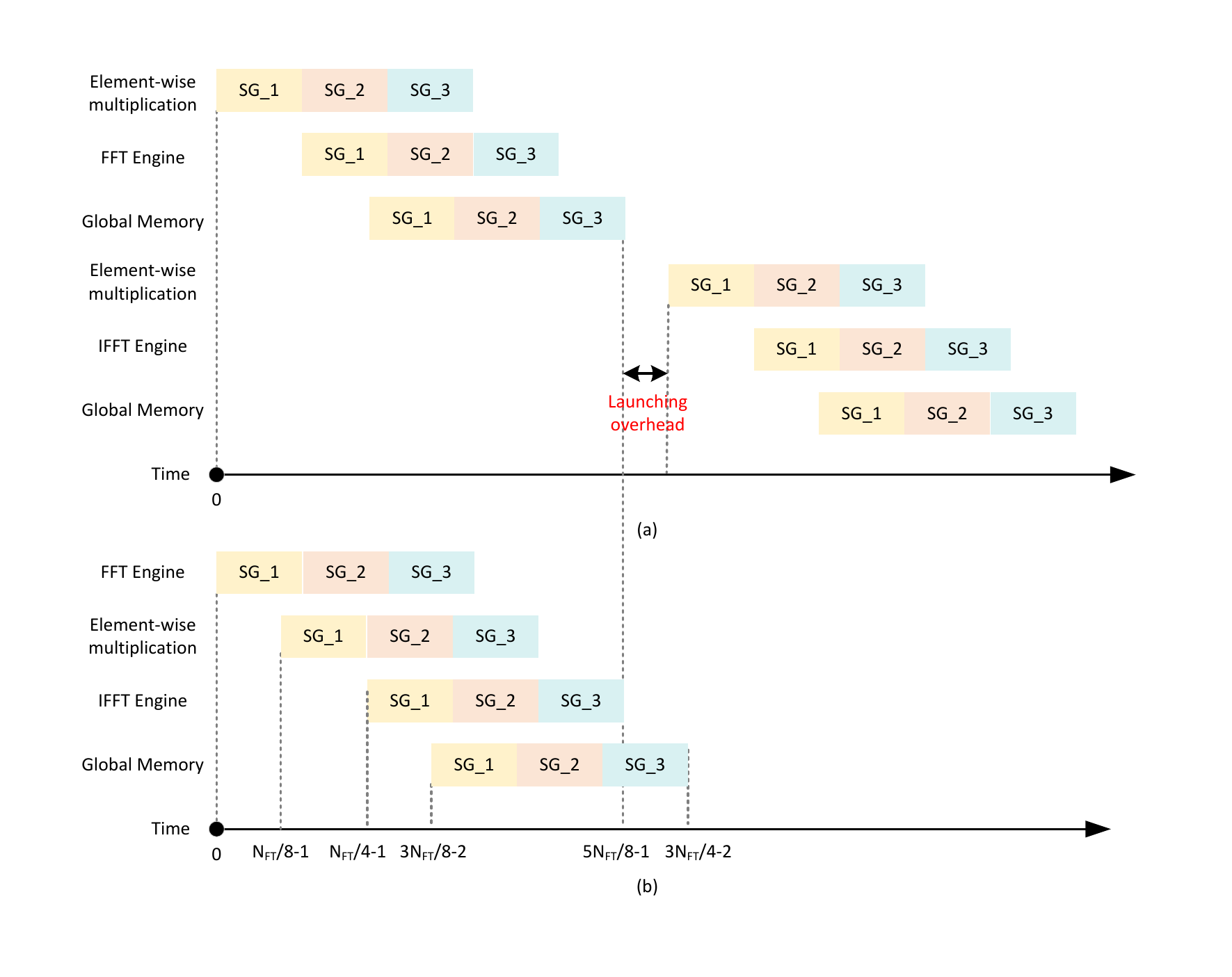}
\par\end{centering}
\caption{\label{fig:Time-cost-of}Time cost of (a)AOLS-FD and (b)TOLS-FD}
\end{figure}

Though the pipeline of TOLS-$N_{FT}$ has more stages than AOLS-$N_{FT}$,
they take about the same amount of clock cycles in one launch when
$N\gg N_{FT}$, which is $\left\lceil \frac{N}{N_{FT}-K}\right\rceil \times\frac{N}{N_{FT-PC}}$.
Since the TOLS-FD structure contains two FFT engines and they can
work simultaneously, it needs fewer clock cycles than AOLS-FD in processing
the same input signals. For example, if the FFT engine is set to process
8 points per clock and the operation frequency of the FPGA is a constant,
then the time costs of using AOLS-FD and TOLS-FD to handle, say, 3
$N_{FT}$ segment arrays are illustrated in Figure~\ref{fig:Time-cost-of}.
Although TOLS-FD structure costs fewer clock cycles, the latency in
practice might not be noticeable, since the operation frequency of
AOLS and TOLS are not the same.

\subsubsection{Multiple FIR Filters\label{subsec:Multiple-FIR-Filters}}

The studied acceleration task for the FT convolution module is to
implement multiple FIR filters, rather than a single FIR filter. We
discussed the implementation of multiple FIR filters as the proposed
structures in Section~\ref{subsec:OLA-TD} and \ref{subsec:OLS-FD}.
Launching an OpenCL kernel on an FPGA (i.e., the kernel has already
been synthesized by AOCL, and the FPGA has been configured correspondingly)
induces a certain time overhead in the order of milliseconds. The
numbers of kernel launches in implementing a single and multiple FIR
filters are given in Table~\ref{tab:Kernel-Launch-Times}.

\begin{table}
\tbl{Kernel Launch Times in Implementing
Single Filter and Multiple Filters\label{tab:Kernel-Launch-Times}}{
\centering{}%
\begin{tabular}{c|ccccc}
\hline 
\multirow{2}{*}{{\scriptsize{}Kernels}} & \multicolumn{2}{c}{{\scriptsize{}TD}} & \multicolumn{3}{c}{{\scriptsize{}FD}}\tabularnewline
 & {\scriptsize{}Na\text{\"i}ve} & {\scriptsize{}OLA} & {\scriptsize{}Na\text{\"i}ve} & {\scriptsize{}AOLS} & {\scriptsize{}TOLS}\tabularnewline
\hline 
{\scriptsize{}Single filter} & {\scriptsize{}$R$} & {\scriptsize{}$R$} & {\scriptsize{}$2$} & {\scriptsize{}$2$} & {\scriptsize{}$1$}\tabularnewline
{\scriptsize{}$M$ filters} & {\scriptsize{}$RM$} & {\scriptsize{}$RM$} & {\scriptsize{}$M+1$} & {\scriptsize{}$M+1$} & {\scriptsize{}$M$}\tabularnewline
\hline 
\end{tabular}}
\end{table}

For TDFIR kernels, each launch takes about $N$ clock cycles. For
OLS-FD kernels, each launch takes about $\left\lceil \frac{N}{N_{FT}-K}\right\rceil \times\frac{N_{FT}}{N_{FT-PC}}$
clock cycles. Different from the OLA-TD kernel, the overlap-add operations
of the Na\text{\"i}ve-TD kernel need to be handled in software by
the host program. For the OLA-TD kernel, its performance is restricted
by the number of parallelised complex SPF multipliers, which is decided
by the available number of DSP blocks, and, to a lesser extent, by
the available amount of the logic resources. To process the same input
array with $M$ FIR filters, the kernel needs to be launched $RM$
times, which is $M$ times implementing single FIR filter. Similar
to the OLA-TD kernel, the available logic resources and DSP blocks
might be a problem for the performance of the TOLS-FD kernel. In implementing
$M$ FIR filters, the TOLS-FD kernel has to be launched $M$ times.
However, the same input array is Fourier transformed $M$ times as
well. 

Regarding the Naive-FD kernel, it executes the 4-million points Fourier
transform and element-wise multiplication during each launch. The
4-million FFT engine, which is provided by Intel, is based on the
general FFT engine and the 4-million points are treated as a $2^{11}\times2^{11}$
matrix. The engine executes the $2K$ FFT on all rows and reorders
the output matrix, followed by another $2K$ FFT on all rows to generate
the Fourier transformed matrix. Different from OLA-TD and TOLS-FD
kernels, the Navie-FD kernel is a generic kernel that can be configured
as FFT or IFFT engine. To implement $M$ FIR filters, the input array
only needs to be Fourier transformed once, and the kernel needs to
be launched $M+1$ times instead of $2M$ times. The AOLS-FD kernel
has the same advantage as Naive-FD, and it needs to be launched $M+1$
times for $M$ FIR filters as well. As the number of FIR filters $M$
increases, the average launching times ($\frac{M+1}{M}$) of each
FIR filter using Naive-FD or AOLS-FD will be halved.

\subsection{Kernel Portability}

OpenCL is designed for developing codes for different target platforms,
however, directly using FPGA-based FIR kernels on other platforms,
such as GPUs, might not achieve high-performance computing. While
there is general portability regarding functionality, performance
portability of FPGA-based FIR filter kernels is affected by three
main factors.

\subsubsection{Single work-item vs. NDRange}

The single work-item Na\text{\"i}ve-TD and OLA-TD kernels include
several optimization techniques specific to the FPGA architecture,
such as unrolling of \texttt{for} loops and shift registers. For GPU-based
implementation, the single work-item kernel will be executed sequentially,
which is similar to a general CPU based implementation. For example,
the performance of single work-item Na\text{\"i}ve kernel in~\cite{wang2015fpga}
(removing FPGA-based optimization code and commands) on a mid-range
AMD GPU is only $0.026GFLOPS$, which is hundreds of times slower
than that of on mid-range FPGA. Regarding the NDRange-based Na\text{\"i}ve-TD
and OLA-TD (Figure~\ref{fig:OpenCL-Code-(using}) kernels, although
the \texttt{for} loop cannot be unrolled by GPU, it can still achieve
high-performance because hundreds to thousands of processing elements
in the GPU can work in parallel. 

\subsubsection{Channel and Pipe}

The AOCL channels are used to connect different kernels in OLS-FD
kernels. Compared with OpenCL pipes, the channel is relatively simple
to use, since it is unnecessary to enable the usage of channels in
the host program. For porting to other platforms, it is better to
use the OpenCL pipe construct, since it conforms to the OpenCL standard.
However, the frequency of channel-based kernels is higher than that
of pipe-based kernels on FPGA devices. Take the TOLS-FD kernel as
an example, it has two FFT engines and all function modules can be
connected using Intel channels or OpenCL pipes. When the $N_{FT}$
is set as 2048, the channel-based kernel frequency is 1.1 times higher
than pipe-based kernels. In our work, the connections between different
kernels are all channels. 

\subsubsection{OpenCL Library}

The employed FFT engine is dedicated for FPGAs since it contains several
FPGA-based optimization techniques and is implemented as a single
work-item kernel. When the employed FPGA-based FFT engine code is
used on GPU or CPU platforms, the performance will be hundreds of
times slower as well. The alternative solution for GPU and CPU platforms
is to use the OpenCL based FFT library called \texttt{clFFT}. Even
though the current AOCL supports OpenCL library technique~\cite{altera2016openclpro},
the \texttt{clFFT} still cannot be used in FPGAs, mainly because it
uses features from OpenCL 1.2, which are not yet supported by AOCL. 

\section{Evaluation}\label{Evaluation}

In this section, we evaluate the proposed FIR filter designs and their
implementations with OpenCL. We do this on two different, but comparable
types of mid-range acceleration devices, namely FPGA and GPU. Our
objective is to determine which of the designs achieves the best performance
and lowest power/energy consumption on FPGAs and put that into relation
to a comparable GPU.

\subsection{Experimental Setup}

The essential characteristics of the employed FPGA and GPU platforms,
both PCIe boards, are given in Table~\ref{tab:Details-of-FPGA}.
For better comparability, the process technology of the Intel Stratix
V 5SGXA7 FPGA and AMD Radeon R7 370 GPU, referred to as $\mathbf{R7}$,
were chosen to be the same, which is $28nm$. 

 For the official provided board support package (BSP) of the DE5
card, the maximum frequency of DDR3 SDRAM is $800MHz$ that makes
the theoretical peak data transfer rate down to $64\times2\times800=102Gbps$.
When the OpenCL kernel frequency $f_{max}$ is smaller than $200MHz$,
the maximum data transfer rate is affected by $f_{max}$, and the
maximum bandwidth is $64\times2\times4\times f_{max}=512f_{max}$,
where factor 4 is the quarter rate, the largest rate supported by
the FPGA soft memory controller. 

The clock frequency of an $A7$ FPGA-based OpenCL kernel is decided
by many factors, and generally around $150-300MHz$, while the clock
of an $R7$ GPU goes up to $985MHz$. It can be seen that the $R7$
GPUs have several advantages over $A7$ FPGAs in operation frequency,
global memory frequency, and global memory bandwidth. For the FIR
filter kernel, the global memory bandwidth is not a barrier, however,
the operating frequency plays an important role for both FPGA and
GPU.

\begin{table}
\tbl{\label{tab:Details-of-FPGA}Details of FPGA and GPU platforms}{
\centering{}{\scriptsize{}}%
\begin{tabular}{c|cc}
\hline 
{\scriptsize{}Device (Board)} & {\scriptsize{}Terasic DE5-Net} & {\scriptsize{}Sapphire Nitro R7 370}\tabularnewline
\hline 
{\scriptsize{}Hardware} & {\scriptsize{}Intel Stratix V 5SGXA7} & {\scriptsize{}AMD Radeon R7 370}\tabularnewline
{\scriptsize{}Technology} & {\scriptsize{}$28nm$} & {\scriptsize{}$28nm$}\tabularnewline
\multirow{2}{*}{{\scriptsize{}Compute resource}} & {\scriptsize{}622,000 LEs} & \multirow{2}{*}{{\scriptsize{}1024 Stream Processors}}\tabularnewline
 & {\scriptsize{}256 DSP blocks} & \tabularnewline
{\scriptsize{}On-chip memory size} & {\scriptsize{}50$Mb$} & {\scriptsize{}\textemdash{}}\tabularnewline
{\scriptsize{}Global memory size} & {\scriptsize{}2 x $2GB$ DDR3} & {\scriptsize{}$4GB$ GDDR5}\tabularnewline
{\scriptsize{}Global memory frequency} & {\scriptsize{}$800MHz$} & {\scriptsize{}$5,600MHz$}\tabularnewline
{\scriptsize{}Memory interface width} & {\scriptsize{}2 x 64-bit} & {\scriptsize{}256-bit}\tabularnewline
{\scriptsize{}Max clock frequency} & {\scriptsize{}\textemdash{}} & {\scriptsize{}985$MHz$}\tabularnewline
{\scriptsize{}OpenCL} & {\scriptsize{}1.0} & {\scriptsize{}1.2}\tabularnewline
{\scriptsize{}Max power consumption} & {\scriptsize{}\textemdash{}} & {\scriptsize{}150W}\tabularnewline
\hline 
\end{tabular}}{\scriptsize \par}
\end{table}

In our evaluation, both FPGA and GPU devices are connected with the
host through 8 lanes (x8) PCIe bus (Gen2.0, $4GB/s$) and the operating
system of the host is Ubuntu 14.04LTS. In terms of the compiler, the
FPGA-based kernels are compiled by AOCL version 15.0.0.145, and GPU-based
kernels are using AMD APP SDK version 3.0~\cite{appsdkgpu,amdappsdk}.

Regarding the measurement made in this research, we measure the execution
latency from starting the FT convolution module in the host program
until the acceleration devices finish processing the input points.
Having that said, input and output points transfer to and from the
host processor is not included in the measurements as the filtering
task is a part of the signal processing pipeline of the FDAS module,
and it is assumed that previous and subsequent modules are also executed
on the acceleration device. Note that, the DE5 card and the graphics
card that employed in this paper are not the final devices for SKA1
CSP PSS deployment. Regarding the proposed structures, when the optimised
approach is specified, it will be implemented using HDLs.

\subsection{FPGA Resource Usage}

\begin{table}
\tbl{\label{tab:Resource-Usage-of}Resource usage of FPGA-based FIR filter
kernels}{
\centering{}{\scriptsize{}}%
\begin{tabular}{cc|cccccc}
\hline 
\multicolumn{2}{c|}{{\scriptsize{}Kernels}} & {\scriptsize{}Logic } & {\scriptsize{}DSP } & {\scriptsize{}RAM } & {\scriptsize{}$f_{max}$ } & {\scriptsize{}Theoretical latency} & {\scriptsize{}rRMSE}\tabularnewline
 &  &  &  &  & {\scriptsize{}($MHz$)} & {\scriptsize{}($ms$)} & {\scriptsize{} ($\times10^{-7}$)}\tabularnewline
\hline 
\multirow{4}{*}{{\scriptsize{}TD}} & {\scriptsize{}TD-Na{\"i}ve-64S} & {\scriptsize{}49\%} & {\scriptsize{}100\%} & {\scriptsize{}15\%} & {\scriptsize{}254.77} & {\scriptsize{}115.24} & \textbf{\scriptsize{}0.695}\tabularnewline
 & {\scriptsize{}TD-Naive-64N} & {\scriptsize{}51\%} & {\scriptsize{}100\%} & {\scriptsize{}18\%} & \textbf{\scriptsize{}270.05} & \textbf{\scriptsize{}108.72} & {\scriptsize{}0.699}\tabularnewline
 & {\scriptsize{}OLA-64S} & {\scriptsize{}50\%} & {\scriptsize{}100\%} & {\scriptsize{}16\%} & {\scriptsize{}236.01} & {\scriptsize{}124.40} & {\scriptsize{}1.94}\tabularnewline
 & {\scriptsize{}OLA-64N} & {\scriptsize{}51\%} & {\scriptsize{}100\%} & {\scriptsize{}20\%} & {\scriptsize{}255.29} & {\scriptsize{}115.01} & {\scriptsize{}1.77}\tabularnewline
\hline 
\multirow{5}{*}{{\scriptsize{}FD}} & {\scriptsize{}FD-Na{\"i}ve} & {\scriptsize{}59\%} & {\scriptsize{}87\%} & {\scriptsize{}90\%} & {\scriptsize{}183.55} & {\scriptsize{}\textendash{}} & {\scriptsize{}2.89}\tabularnewline
 & {\scriptsize{}AOLS-1024} & {\scriptsize{}52\%} & {\scriptsize{}50\%} & {\scriptsize{}46\%} & \textbf{\scriptsize{}222.17} & {\scriptsize{}7.98} & \textbf{\scriptsize{}1.69}\tabularnewline
 & {\scriptsize{}AOLS-2048} & {\scriptsize{}54\%} & {\scriptsize{}59\%} & {\scriptsize{}88\%} & {\scriptsize{}205.59} & {\scriptsize{}6.41} & {\scriptsize{}1.78}\tabularnewline
 & {\scriptsize{}AOLS-4096} & {\scriptsize{}59\%} & {\scriptsize{}60\%} & {\scriptsize{}72\%} & {\scriptsize{}173.97} & {\scriptsize{}6.72} & {\scriptsize{}1.86}\tabularnewline
 & {\scriptsize{}TOLS-1024} & {\scriptsize{}83\%} & {\scriptsize{}88\%} & {\scriptsize{}77\%} & {\scriptsize{}168.26} & \textbf{\scriptsize{}5.27} & {\scriptsize{}3.26}\tabularnewline
\hline 
\end{tabular}}{\scriptsize \par}
\end{table}

Before evaluating the execution latency and the energy dissipation
that are discussed in the next section, we focus on the FPGA resource
consumption and the correctness. Table~\ref{tab:Resource-Usage-of}
lists the resource usage, maximum kernel frequency, and the theoretical
latency of all proposed FIR filter kernels. The latency is calculated
based on processing $2^{22}$ complex SPF points using a single 421-tap
FIR filter. To confirm the correctness of the SPF outputs, we calculate
the relative Root Mean Square Error (rRMSE) of the FPGA output points
in comparison to the Matlab results, whose data type is double precision
floating-point. For the TDFIR kernels, the 'S' and 'N' after the parallelisation
factor 64 represent the \textbf{S}ingle work-item kernel and \textbf{N}DRange
kernel types. For all FDFIR kernels in Table~\ref{tab:Resource-Usage-of},
the FFT engine processes 8 points per clock cycle. 

For the theoretical latencies, they are calculated based on the required
clock cycle (based on the values in Table~\ref{tab:Workload-of-one}
and Table~\ref{tab:Kernel-Launch-Times}) and kernel frequency $f_{max}$
(in Table~\ref{tab:Resource-Usage-of}). The kernel launching overhead
is unpredictable and it is not included in the theoretical latencies.
Due to the substantial difference in the approaches, the performances
regarding $GFLOPS$ values of the TDFIR kernels are not comparable
with those of the FDFIR kernels. For OLS-FD kernels, part of the points
from each $N_{FT}$ Fourier transformed values have to be discarded
(due to the overlap). Hence, the real performance of these kernels
is higher than the valid performance. Take AOLS-1024 as an example,
over 40\% of each 1,024 Fourier transformed points are discarded,
which means the real performance is up to 1.6x times higher than the
valid performance in Table~\ref{tab:Resource-Usage-of}. 

All TDFIR kernels are configured to the maximum size as they use 100\%
of the available DSP blocks, which is the same as theoretical analysis.
For FDFIR kernels, the DSP blocks usage does not reach 100\%, and
the resource usage is balanced. The real DSP blocks usage of AOLS-$N_{FT}$
and TOLS-$N_{FT}$ is the same as discussed in Section~\ref{subsec:Optimisation-for-FT}.
As one can expect the TOLS variant uses significantly more resources
than the comparable (ALOS-1024) variant, where the DSP usage is about
1.75x times higher.

It can be seen that the theoretical latency of Na\text{\"i}ve-64N
is the best among TDFIR kernels, however, it can only execute one
FIR filter that is up to 64 taps long. Positive is that the flexible
OLA-64N filter has very similar theoretical latency. In terms of the
FDFIR kernel, TOLS-1024 performs better than other FDFIR kernels,
and AOLS-2048 performs better than other AOLS-FD with different $N_{FT}$.
However, the advantage of kernel TOLS-1024 over AOLS-FD kernels might
disappear when implementing multiple FIR filters (see Section~\ref{subsec:Multiple-FIR-Filters}).

Regarding the kernels that calculate the power of complex value, we
implemented six different combinations of AOLS-$N_{FT}$-P kernels
employing a 4-point FFT engine. The details of resource usage, maximum
frequency, and theoretical latency are provided in Table \ref{tab:TDFIR-Output-Power},
where the theoretical latency is the average latency for a single
FIR filter. Based on the theoretical latency, kernel AOLS-2048-P performs
better than all kernels in Table~\ref{tab:Resource-Usage-of}, when
the kernel is replicated three times. For all configurations, where
the structure is replicated three times, the DSP block usages are
over 80\%, and especially for 3xAOLS-4096-P, most resources on an
$A7$ FPGA reach exhaustion. Because of the high percentage of resource
usage, the achieved kernel frequency is lower than other kernels.

\begin{table}
\tbl{\label{tab:TDFIR-Output-Power}Resource usage of power calculation
kernels (P represents that output is power of complex numbers)}{
\centering{}{\scriptsize{}}%
\begin{tabular}{cc|ccccc}
\hline 
{\scriptsize{}Kernels} & {\scriptsize{}Number} & {\scriptsize{}Logic } & {\scriptsize{}DSP } & {\scriptsize{}RAM } & {\scriptsize{}$f_{max}$} & {\scriptsize{}Theoretical latency}\tabularnewline
 &  & {\scriptsize{}utilization} & {\scriptsize{}blocks} & {\scriptsize{}blocks} & {\scriptsize{}($MHz$)} & {\scriptsize{}($ms$)}\tabularnewline
\hline 
{\scriptsize{}AOLS-1024-P} & {\scriptsize{}2} & {\scriptsize{}54\%} & {\scriptsize{}57\%} & {\scriptsize{}37\%} & {\scriptsize{}216.35} & {\scriptsize{}4.12}\tabularnewline
{\scriptsize{}AOLS-1024-P} & {\scriptsize{}3} & {\scriptsize{}71\%} & {\scriptsize{}85\%} & {\scriptsize{}54\%} & {\scriptsize{}214.27} & {\scriptsize{}2.77}\tabularnewline
\hline 
{\scriptsize{}AOLS-2048-P} & {\scriptsize{}2} & {\scriptsize{}58\%} & {\scriptsize{}66\%} & {\scriptsize{}59\%} & \textbf{\scriptsize{}220.21} & {\scriptsize{}3.00}\tabularnewline
{\scriptsize{}AOLS-2048-P} & {\scriptsize{}3} & {\scriptsize{}68\%} & {\scriptsize{}98\%} & {\scriptsize{}81\%} & {\scriptsize{}205.68} & \textbf{\scriptsize{}2.14}\tabularnewline
\hline 
{\scriptsize{}AOLS-4096-P} & {\scriptsize{}2} & {\scriptsize{}59\%} & {\scriptsize{}66\%} & {\scriptsize{}92\%} & {\scriptsize{}182.68} & {\scriptsize{}4.56}\tabularnewline
{\scriptsize{}AOLS-4096-P} & {\scriptsize{}3} & {\scriptsize{}72\%} & {\scriptsize{}99\%} & {\scriptsize{}88\%} & {\scriptsize{}177.24} & {\scriptsize{}3.13}\tabularnewline
\hline 
\end{tabular}}{\scriptsize \par}
\end{table}

\subsection{Performance Comparison}

In this section, we now evaluate and compare the execution time or
latency of processing the entire input completely. While we provide
performance values regarding $GFLOPS$ for the TDFIR implementations,
comparing this value between TDFIR and FDFIR kernels would be misleading,
as the algorithms and necessary computations are significantly different,
not incurring the same number of operations to perform the filtering.

\subsubsection{TDFIR vs FDFIR}

\begin{figure}
\begin{centering}
\includegraphics[bb=0bp 10bp 490bp 310bp,clip,scale=0.55]{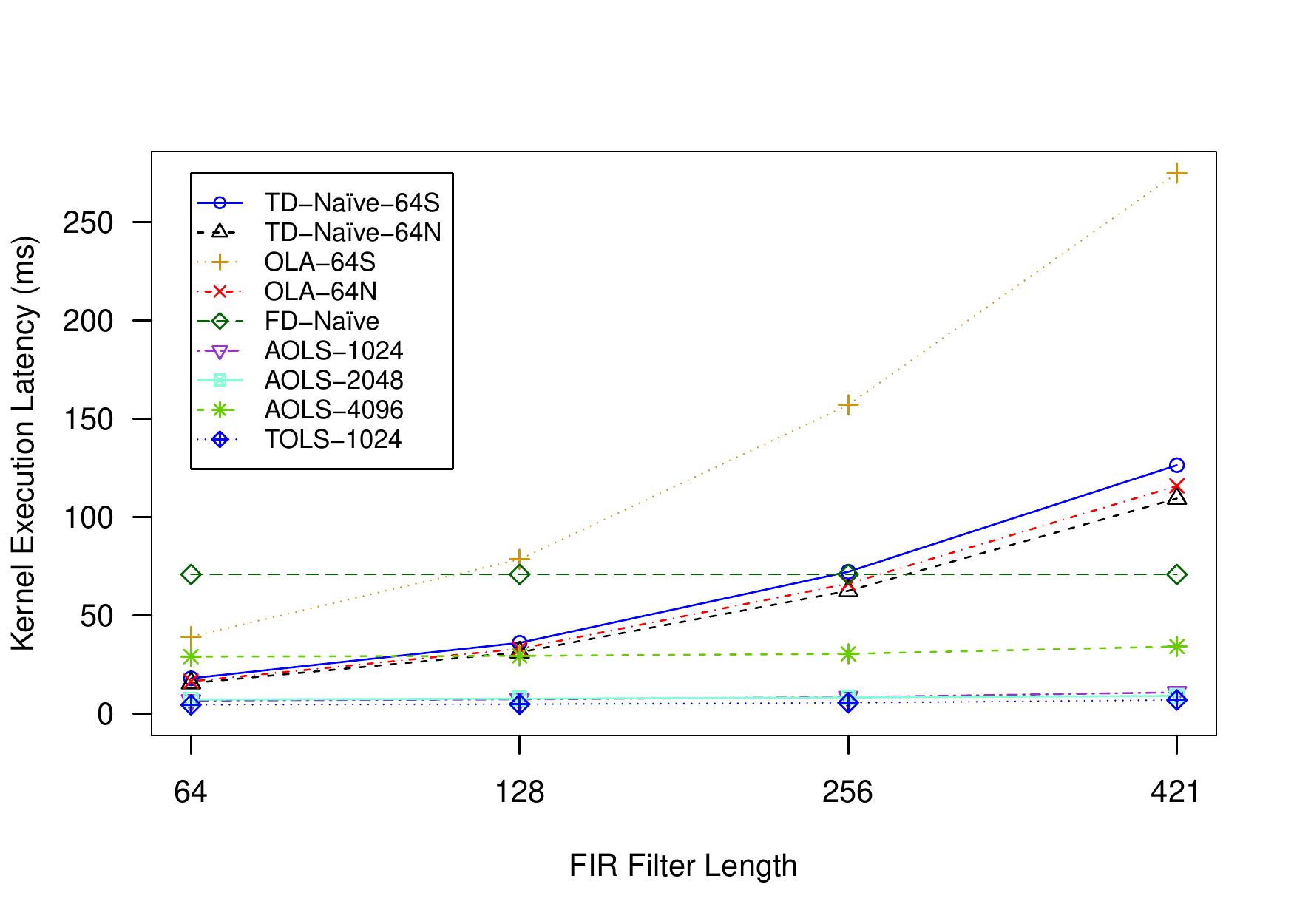}
\par\end{centering}

\protect\caption{\label{fig:Kernel-Execution-Latencies}Kernel execution latencies
of FPGA-based FIR filter kernels}
\end{figure}

The execution latencies of all nine kernels in Table~\ref{tab:Resource-Usage-of}
are plotted in Figure~\ref{fig:Kernel-Execution-Latencies} over
different filter lengths. All these kernels are launched to process
4 million complex SPF numbers using 64, 128, 256, and 421-tap FIR
filters, respectively. It can be seen that TOLS-1024 is the fastest
of all nine kernels and AOLS-2048 is the second when implementing
a 421-tap FIR filter. For TDFIR kernels, additional operations are
needed to accumulate the results by the host when using kernel Na\text{\"i}ve-64S
and kernel Na\text{\"i}ve-64N to implement an FIR filter larger than
64 taps. With this in mind, kernel OLA-64N is the best of 4 TDFIR
kernels in implementing large FIR filter. The FIR filter length does
not affect the performance of FDFIR kernels too much, while the latencies
of all TDFIR kernels are raised steadily as the filter size increases.
In implementing a single 421-tap FIR filter, all FDFIR kernels perform
better than the TDFIR kernels, so we focus on FDFIR-based kernels
now. 

When comparing the real results with the theoretical latencies in
Table~\ref{tab:Resource-Usage-of}, the real results of NDRange-based
TDFIR kernels is about the same as the theoretical latencies (1.01x
times slower). Regarding the OLS-FD kernels, the actual results are
over 1.3x times slower than the theoretical latencies. The main reason
is that the kernel launching overhead is of the same order of magnitude
with the execution latency. AOLS-2048 performs better than AOLS-1024
because the proportion of invalid points per $N_{FT}$ points of AOLS-2048
is smaller than that of AOLS-1024. For both AOLS-2048 and AOLS-1024,
the points can be streamed between the off-chip memory and the FPGA.
Regarding AOLS-4096, it performs worse than AOLS-2048 and much worse
than the estimated latency. The distances between addresses of necessary
points to Fourier transfer 4096 points are too large, which makes
AOLS-4096 fails to achieve the streaming mode.

\subsubsection{FDFIR with Spectral Power Calculation}

\begin{figure}
\begin{centering}
\includegraphics[bb=0bp 0bp 490bp 330bp,clip,scale=0.55]{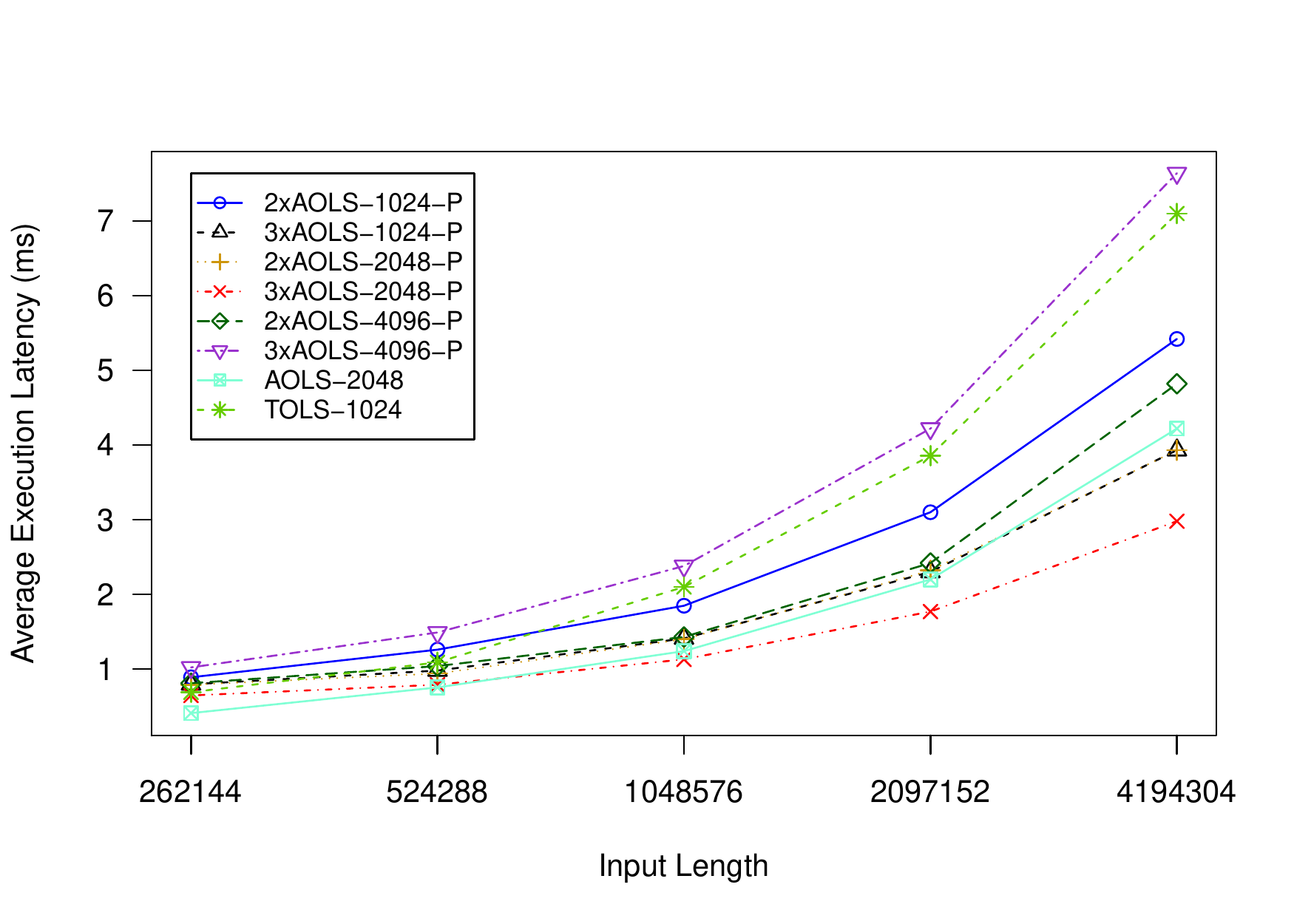}
\par\end{centering}
\caption{\label{fig:Kernel-power-Execution-Latencies}Average execution
latencies of two FDFIR kernels and six (multiple) kernels with appended
power calculation}
\end{figure}

All kernels in Table~\ref{tab:TDFIR-Output-Power} are evaluated
and compared with two pure FDFIR kernels (TOLS-1024 and AOLS-2048
as of Figure~\ref{fig:Kernel-Execution-Latencies}). All these kernels
are employed to implement 84 different 421-tap FIR filters, and the
range of input array size is from $2^{18}$ to $2^{22}$. The average
execution latency of one 421-tap FIR filter is plotted in Figure~\ref{fig:Kernel-power-Execution-Latencies}.
It can be seen that kernel AOLS-2048 performs better than kernel TOLS-1024
in implementing multiple FIR filters, which stems largely from the
fact that the forward FFT only needs to be executed once with AOLS-2048.
The average latencies of four kernel configurations with appended
power calculation are smaller than that of pure AOLS-2048, namely
3xAOLS-1024-P, 2xAOLS-2048-P, 3xAOLS-2048-P, and 2xAOLS-4096-P. All
these four kernels not only process additional operations (calculate
the power of each complex value) but also perform better than all
generic FIR kernels in implementing multiple FIR filters. Kernel 3xAOLS-2048-P
performs best among these eight kernels, and the average latency of
processing $2^{22}$ points is $2.98ms$, which is the same as predicted
in Table~\ref{tab:TDFIR-Output-Power}. In total, it takes over $250ms$
to apply 84 FIR filters. 

\subsubsection{Multiple FPGAs}

\begin{figure}
\begin{centering}
\includegraphics[scale=0.55]{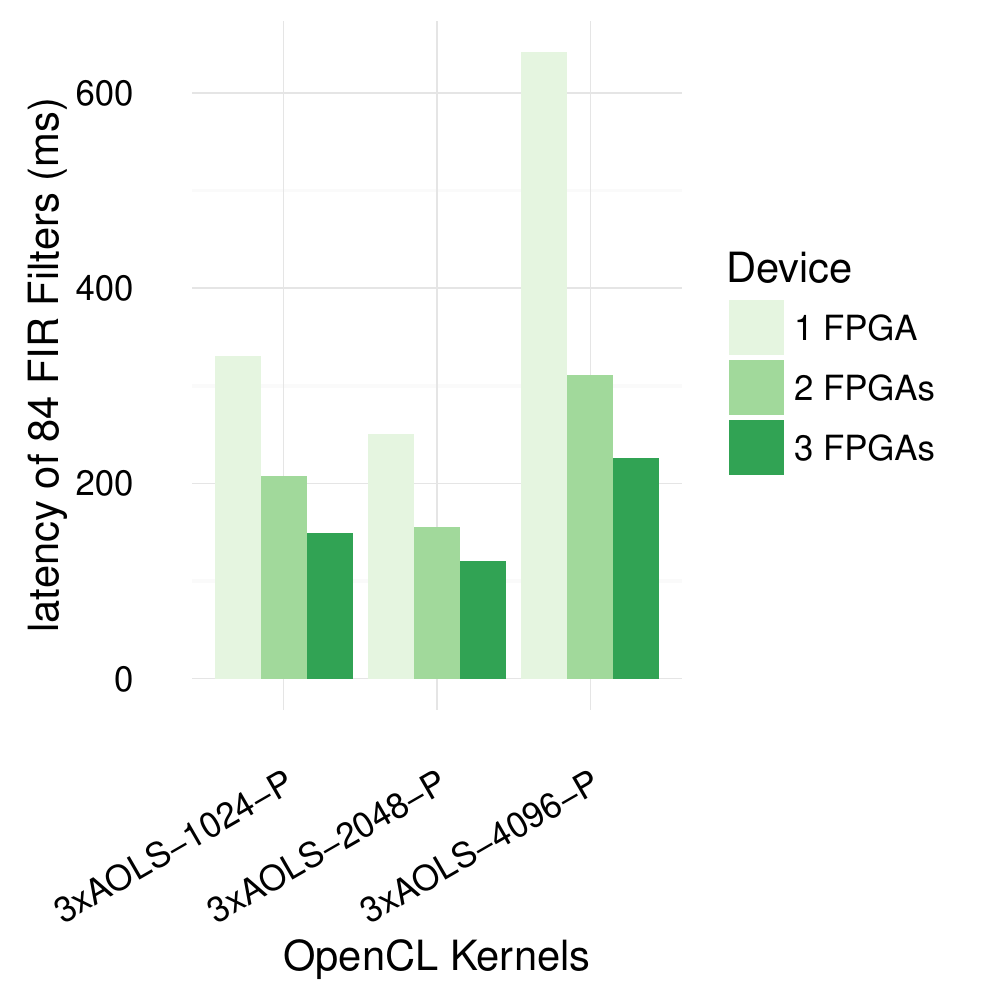}\includegraphics[scale=0.55]{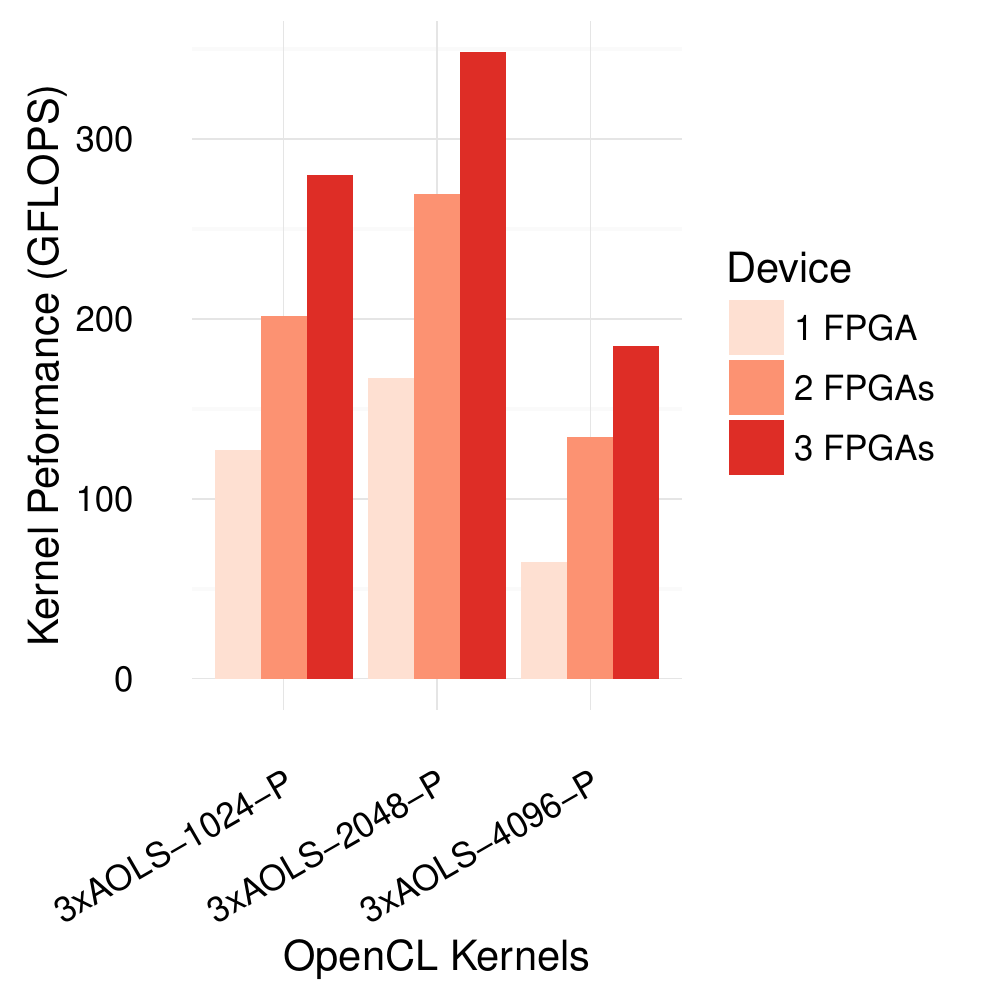}
\par\end{centering}
\caption{\label{fig:OpenCL-kernels-on}Latency and performance of OpenCL kernels
using multiple FPGA devices}
\end{figure}

While intensively exploiting the FPGA resources and memory bandwidth,
$250ms$ is still significantly larger than the time limit of the
FT convolution module which is under $100ms$. Thus we investigate
using multiple FPGA devices to reduce the latency. Up to three FPGA
devices (DE5-Net boards) are used in our work, and the latency and
performance are plotted in Figure~\ref{fig:OpenCL-kernels-on}. We
compare the three kernels with three instances each from the previous
experiment, i.e., 3xAOLS-1024-P, 3xAOLS-2048-P, 3xAOLS-4096-P varying
the number of FPGAs (boards) used in one single host system. As in
the single FPGA case, 3xAOLS-2048-P performs better than the other
two kernels on multiple FPGA devices. Using three FPGA devices, kernel
3xAOLS-2048-P can apply 84 FIR filters in $120ms$, and the effective
performance is around $350GFLOPS$.

\subsubsection{FPGA vs GPU}

Now the FPGA kernels are compared to a GPU implementation. The GPU
operating frequency is locked at the maximum frequency, which is $985MHz$,
and all 16 compute units are set to process in parallel. Since the
$R7$ GPU has over 1,000 stream processors, there is no need to use
OLA algorithm on it. Regarding FDFIR, it takes tens milliseconds for
$R7$ GPU to process a 4-million FFT, so the OLS algorithm is employed
for GPU. 

The single work-item kernels are unfair for GPUs, which takes several
seconds to execute, so we only compare the NDRange kernels on FPGA
and GPU. Two FPGA-based kernels are evaluated on $R7$ GPU: NDRange
TD-Na\text{\"i}ve and AOLS-2048. The FPGA-based NDRange TD-Na\text{\"i}ve
kernel can be ported directly to the $R7$ GPU, referred to as \textbf{GPU-TD}.
For GPU-based AOLS-2048, referred to as \textbf{GPU-FD}, the FFT engine,
which is a single work-item kernel, is replaced with \texttt{clFFT}.
The \texttt{clFFT} is an API designed for AMD's graphics card to perform
FFT, which is well-optimised for GPU.

\begin{figure}
\begin{centering}
\includegraphics[bb=0bp 10bp 504bp 250bp,clip,scale=0.55]{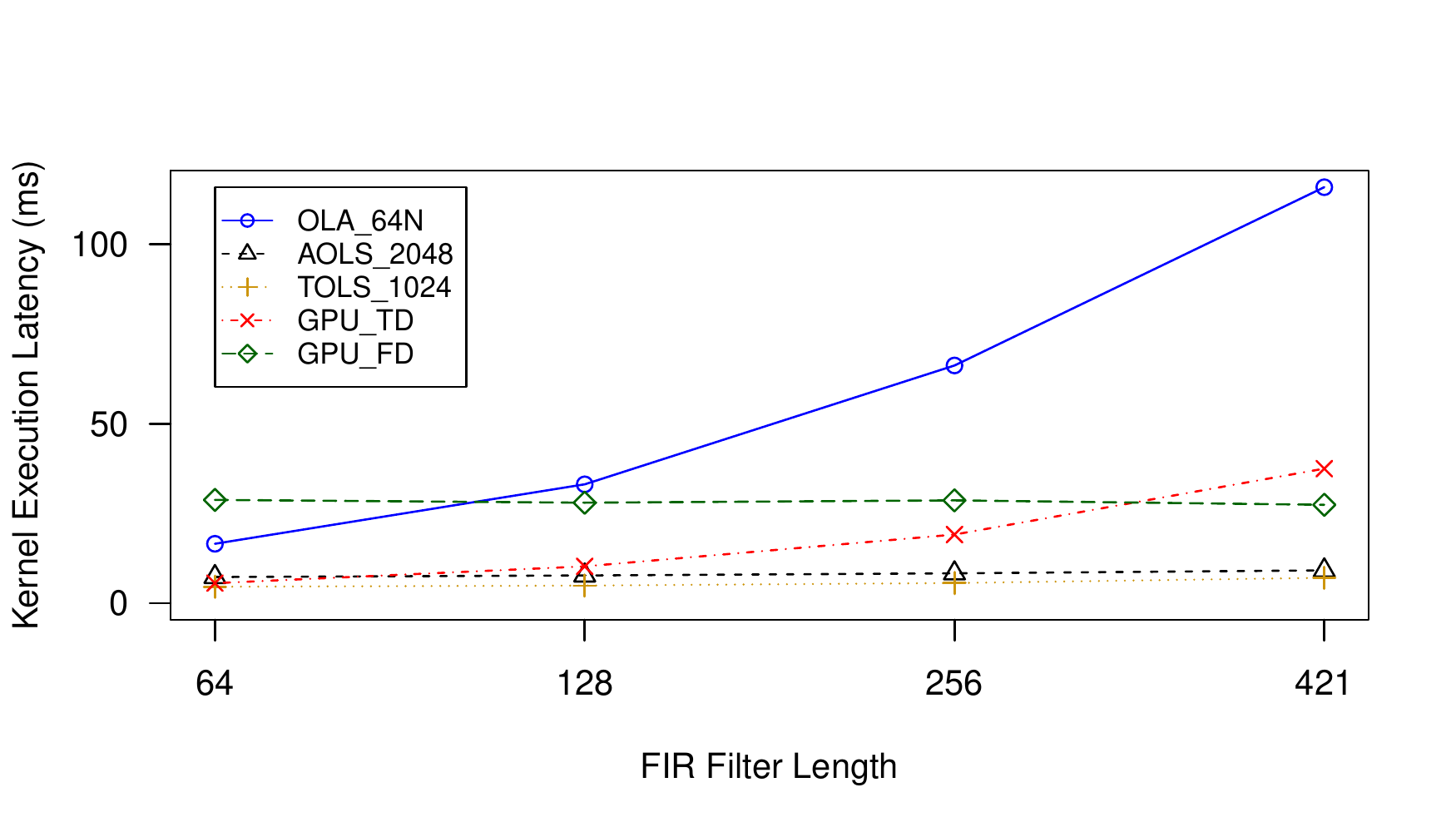}
\par\end{centering}
\caption{\label{fig:Latency-of-GPU}Execution latencies of GPU and FPGA-based
kernels in implementing a single FIR Filter}
\end{figure}

Now to the actual kernel performance comparison, we compared the execution
latency of the fastest FPGA-based TDFIR and FDFIR kernels, which are
OLA-64N, AOLS-2048, and TOLS-1024, with GPU-TD and GPU-FD. The latencies
of these kernels in implementing a single 421-tap FIR filter are plotted
in Figure~\ref{fig:Latency-of-GPU} over different filter lengths. 

We observe that as the FIR filter length increases, the trends of
GPU-based kernels are similar to those of FPGA-based kernels. However,
the performance of GPU-TD kernel is over three times higher than the
fastest $A7$ FPGA-based TDFIR kernel, which is over $\mathbf{450GFLOPS}$
when implementing a 421-tap FIR filter. This is caused by the high
operating frequency of GPU device. In terms of the GPU-FD kernel,
the performance of it is mainly decided by the operation frequency
of the GPU and not by the FIR filter length. For a single FIR filter,
the GPU-FD kernel performs worse than the two FPGA-based FDFIR kernels.
However, in comparison with 1D convolution results in~\cite{fowers2013performance},
the GPU-FD performs much better than the best solution of it, where
the data type is SPF instead of complex SPF.

\begin{figure}
\begin{centering}
\includegraphics[bb=0bp 10bp 504bp 350bp,clip,scale=0.6]{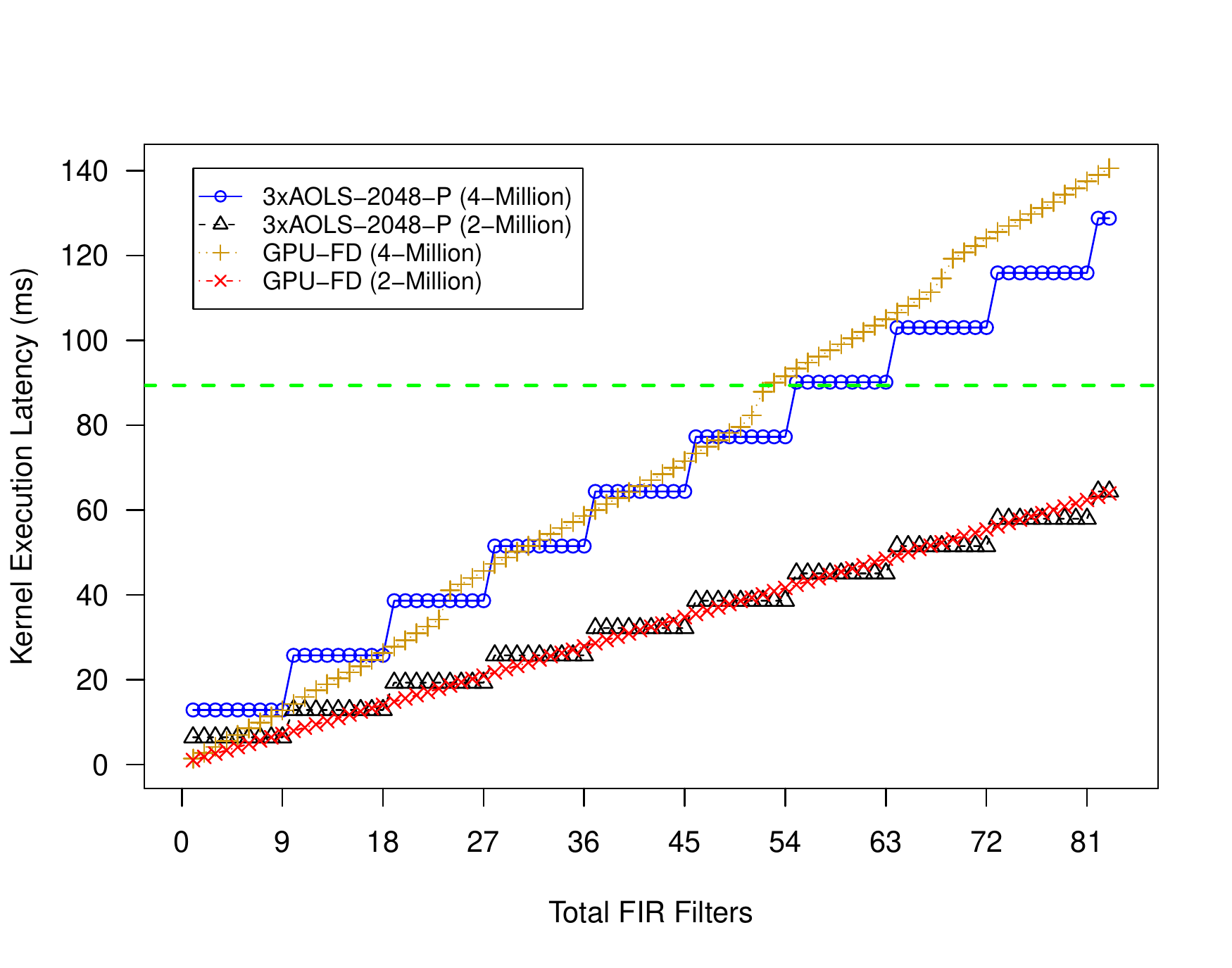}
\par\end{centering}
\caption{\label{fig:Latency-of-Executing}Execution latencies of GPU and FPGA
based kernels in implementing multiple FIR filters}
\end{figure}

For applying multiple FIR filters, we compare the best performance
we achieved using multiple FPGA devices (3 FPGAs with 3xAOLS-2048-P)
with a single $R7$ GPU, whose execution latencies are charted in
Figure~\ref{fig:Latency-of-Executing}. Two different input sizes
are evaluated, which have 2 million and 4 million points. Three FPGA-3xAOLS-2048-P
implement 9 FIR filters during each launch. So the latency of executing
a single FIR filter has no difference with that of 9 FIR filters.
This is the reason for the steps that can be observed in the curves
in Figure~\ref{fig:Latency-of-Executing}. The execution latency
of a single GPU is about the same as that of three FPGA devices in
processing both 2-million and 4-million points. For 4-million points,
the 3xAOLS-2048-P kernel on the three $A7$ FPGAs performs relatively
better when the total number of FIR filters is larger than 64. Also
remember that the 3xAOLS-2048-P kernel performs more computations
than the GPU-FD, as the calculation of the power value of each complex
point is included.

It should be noted that none of the considered implementations can
finish processing 4-million points in the specified time limitation
of $89.4ms$ (green dot line). If the input size can be reduced to
2-Million points or the number of FIR filters can be reduced to 53
or less, it is possible for a single $R7$ GPU or three $A7$ FPGAs
to handle the FT convolution module for one beam. 

\subsection{Power and Energy Consumption}

In the previous section, we saw that the execution time performance
of three $A7$ FPGAs and one $R7$ GPU is very similar. It is now
very interesting to compare their power consumption.

\subsubsection{Power Measurment}

Both FPGA and GPU devices are employed as acceleration hardware in
our research. To compare the power consumption of such computation,
we need to take the power consumption of both device and host into
consideration. 

A plug-in power meter (Efergy Ego Smart Power Socket) is used to measure
the power consumption of the overall system. When the acceleration
devices, both FPGA and GPU, are idle, they still need additional power,
referred to as $P_{device-idle}$, especially for FPGA devices. After
configuring the FPGA, the value of $P_{device-idle}$ for different
kernels are different. For DE5 board, it costs $10-20W$ without executing
any tasks and $10-15W$ for an $R7$ GPU board.

There are two main steps to measure the power consumption of an acceleration
device based computing. First, before installing acceleration devices,
the power consumption of the basic host in the idle state is measured,
referred to as $P_{host-idle}$. Then we install the devices and launch
a kernel for up to 5 minutes by using\texttt{ }loops till the measured
power consumption in watt becomes stable. The constant power consumption
of the running system is recorded as $P_{total}$. The real power
consumption of $Kernel_{i}$ can be calculated as 
\[
P_{Kernel_{i}}=P_{total}-P_{host-idle}.
\]
The value of $P_{Kernel_{i}}$ is not only the power cost of the devices
but the overall cost of using acceleration devices to process a task.
It consists of two parts: the power consumption of the acceleration
devices and the power consumption of the host in setting kernel arguments
and launching kernels. However, the power consumption of acceleration
devices is the largest of these three parts.

\subsubsection{Power Comparison}

\begin{figure}
\begin{centering}
\includegraphics[bb=0bp 0bp 288bp 270bp,clip,scale=0.55]{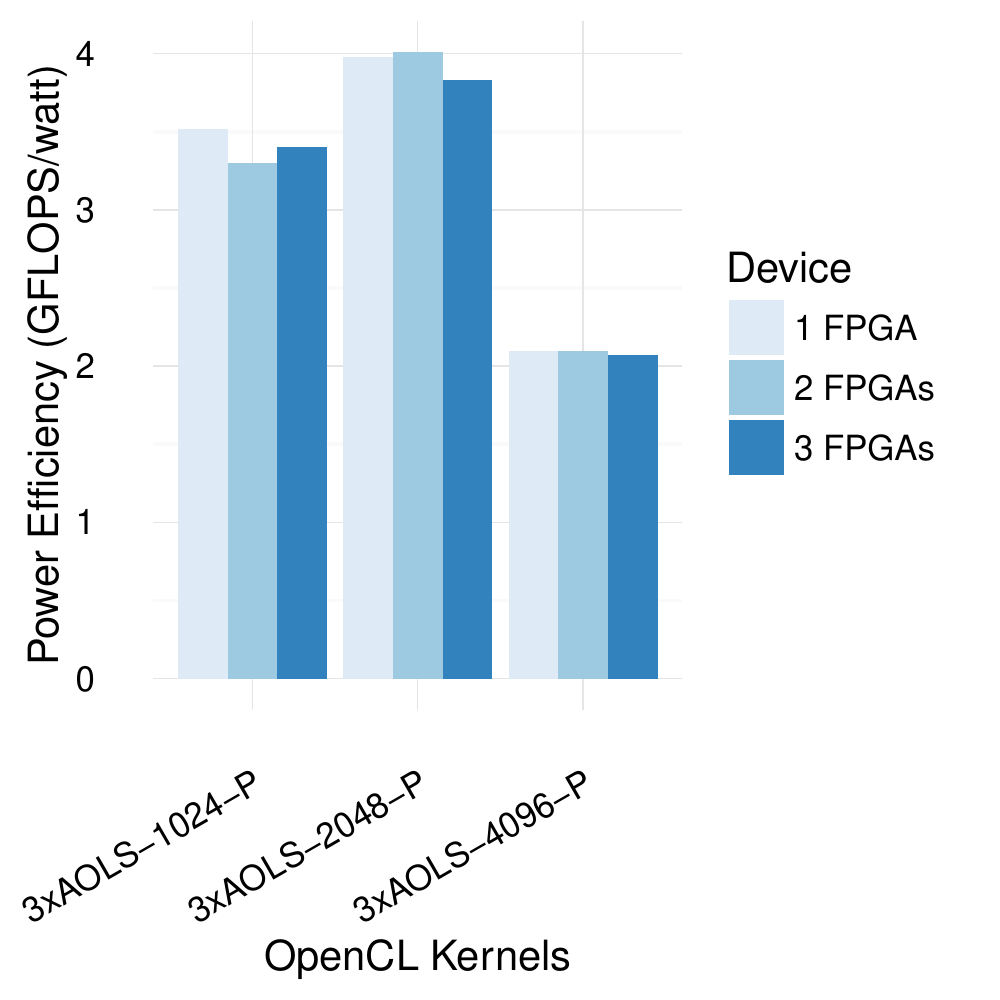}\includegraphics[bb=0bp 0bp 288bp 270bp,clip,scale=0.55]{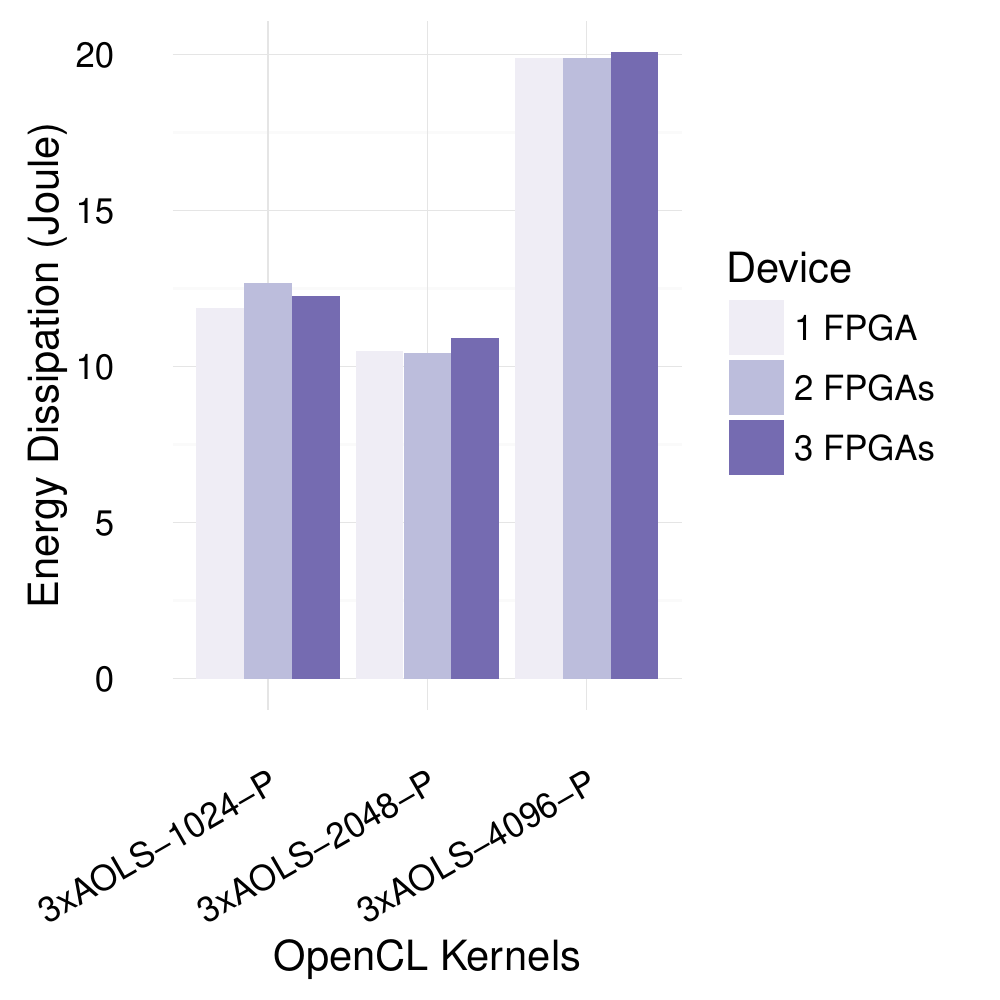}
\par\end{centering}
\caption{\label{fig:power fpga}Power efficiency and energy dissipation of
executing multiple FIR filters}
\end{figure}

The power efficiency and energy dissipation of multiple FPGA-based
AOLS-$N_{FT}$-P kernels are depicted in Figure~\ref{fig:power fpga}.
It can be noted that the number of FPGA devices does not influence
the power efficiency and energy dissipation of AOLS-$N_{FT}$-P kernels
too much. For all three AOLS-$N_{FT}$-P kernels, the power efficiency
remains stable as the number of FPGA devices increases. For a specified
task, the energy dissipation of it is decided by the power efficiency
of the kernels. When the workload of the task is fixed, the higher
the value of power efficiency of a kernel, the less energy it dissipates.
So we mainly investigate the energy dissipation of single device based
kernels. 

\begin{figure}
\begin{centering}
\includegraphics[bb=0bp 0bp 504bp 330bp,clip,scale=0.55]{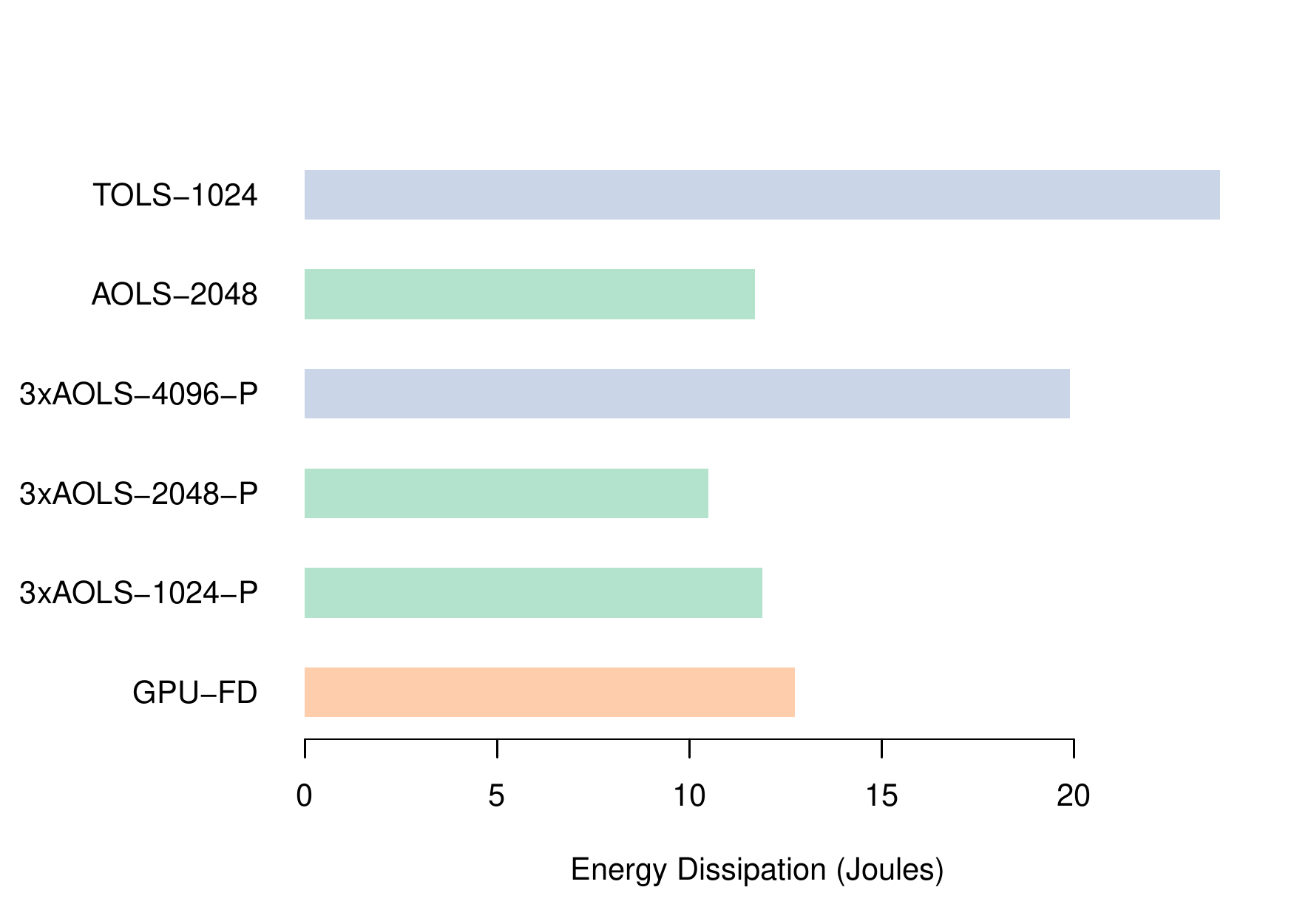}
\par\end{centering}
\caption{\label{fig:power comparison}Energy consumption of executing multiple
FIR filters}
\end{figure}

The average power consumption of running an $R7$ GPU device is $90-105W$
and higher than that of a single $A7$ FPGA device, which ranges from
$20W$ to $40W$. The energy dissipation of five high-performance
FPGA-based kernels is compared with the GPU-based GPU-FD kernel in
processing 4-million points with 84 different 421-tap FIR filters,
which is shown in Figure~\ref{fig:power comparison}. The energy
dissipation of kernel AOLS-2048 and 3xAOLS-1024-P is both fewer than
that of kernel GPU-FD, even though the performance is worse than that
of GPU-FD. Kernel 3xAOLS-2048-P has advantages over kernel GPU-FD
in both performance and energy dissipation, and the energy dissipation
of it the fewest among all evaluated kernels. 

The power needed for three DE5 boards based 3xAOLS-2048-P is about
the same as a single $R7$ GPU based GPU-FD, which is $91W$. However,
each DE5 board has an individual power module and cooling system.
If multiple FPGAs can be integrated into one board, the power cost
might drop, and the advantages of FPGAs over GPUs would further increase.
In processing the same input signals, the FPGA acceleration cards
costs less energy than the GPU card, while providing similar execution
performance. The extremely large-scale nature of the SKA and its longevity
of many years make this an essential advantage of FPGA based solution. 

\section{Conclusion}\label{Conclusion}

This paper investigated the FPGA-based acceleration of the FT convolution
for Pulsar Search in the SKA project using OpenCL as a high-level
development technique. Because of the limitation of memory bandwidth
and resources of an FPGA, the OLA and OLS algorithms were investigated
to make it possible for an FPGA to implement multiple large FIR filters
and process large size input signals. Different approaches to implementing
TDFIR and FDFIR were designed and experimentally evaluated. The results
given evidence that OpenCL can well be used to development FPGA solutions
efficiently while achieving high performance for such computations.
The FPGA-based FDFIR kernels perform better than TDFIR kernels for
lengthy FIR filters. Even though the achieved GFLOPS are higher for
TDFIR, the FDFIR computation is more efficient. We studied different
designs and configurations of the proposed filters to exploit the
available FPGA resources as much as possible. To evaluate the portability
of FPGA-based kernels, and to put the FPGA performance into relation
with GPUS, two appropriate kernels were tested on a mid-range GPU
device. The experiments demonstrate that the latency of FPGA-based
FDFIR kernels is smaller than that of GPU-based kernels for a single
FIR filter. We also investigated the use of multiple FPGA devices
and the computation of multiple filters. Three $A7$ FPGA devices
perform better than single $R7$ GPU device while being more power
efficient.

\begin{acks}
The authors acknowledge discussions with the TDT, a collaboration
between Manchester and Oxford Universities, and MPIfR Bonn and the
work benefitted from their collaboration. 
We gratefully acknowledge that this research was financially supported by the 
SKA funding of the New Zealand government through the Ministry of Business, 
Innovation and Employment (MBIE).
\end{acks}


\end{document}